\documentclass[a4paper,10pt]{article}
\usepackage{ucs}
\usepackage[utf8]{inputenc}
\usepackage{amsmath}
\usepackage{amsfonts}
\usepackage{amssymb}
\usepackage{amsthm}
\usepackage{rotating}
\usepackage{xcolor}
\usepackage[english]{babel}
\usepackage[T1]{fontenc}
\usepackage{graphicx}
\usepackage{hyperref}
\usepackage{cuted}
\usepackage{verbatim}
\usepackage{subfigure}
\usepackage{multirow}
\usepackage{bbold}
\usepackage{slashed}
\usepackage{indentfirst}
\usepackage{tcolorbox}
\usepackage{placeins}
\usepackage[a4paper, total={7in, 10in}]{geometry}
\usepackage{setspace} 
\newcounter{eq}

\newcommand{\bpsi}{\bar{\psi}}

\begin{document}

%\doublespacing

\title{\bf 
Charged pions asymmetry due to interference from the
decay of light neutral axial mesons 
} 

\author{ Fabio L. Braghin \thanks{ braghin@ufg.br}
\\
{\normalsize Instituto de F\'\i sica, Federal University of Goias}
\\
{\normalsize Av. Esperan\c ca, s/n,
 74690-900, Goi\^ania, GO, Brazil }}

\date{}

\maketitle

\begin{abstract}
Effective three meson couplings, based in flavor U(3) quark-antiquark
 interactions,
 are considered to describe  two
light neutral axial meson
Strong decays, 
 $A \sim  f_1(1285)$ and $f_{1S} (1420)$,
and the   charged rho meson decay
in the following channels:
$A \to \rho^\mp (770)+ \pi^\pm$,
 $A\to a_0^\mp (980)  +\pi^\pm$ and
$\rho^\pm \to \pi^\pm \pi^0$.
By considering neutral meson mixings, such as
$f_1(1285) - a_1^0(1260)$, $f_{1S}(1420) - a_1^0(1260)$ and
$\eta-\pi^0$ for  the channels investigated,
the leading  three-meson interactions may lead to 
 decay amplitudes that undergo
 interference and a resulting 
ratio of decay rate into  charged pions, $\Gamma_{\pi^+}/\Gamma_{\pi^-}$,
slightly smaller than one. 
For that it will be assumed 
 that the produced charged vector mesons or charged scalar mesons  
undergo absorption, decay into  channels without charged pions,
 or any other inelastic process that may suppress their decays into charged pions. 
\end{abstract}

\section{ Introduction}

Effective three-meson interactions may
describe particular channels of meson (strong) decay  into two mesons 
and several types 
 have been  derived and investigated with different approaches
\cite{witten,NJL2,EffMod,several-Seff,hiller-etal,PRD-2022b,FLB-2024b}.
 These couplings may  correspond to  a  mechanism for meson production  
in  (relativistic) heavy ion collisions,
usually based in quark polarization processes.
Meson production has been  extensively investigated in the last decades,
being their  description  usually   difficult since they may have associated
 to final state  strong  interactions
\cite{partprod1,partprod2,Floris,FOPI,Zhang-Rho,JunXu-etal}.
Several facilities and experiments  investigate and plan to investigate related aspects in a wide 
regime of energies,
that may involve meson production, polarization and their decays, such as
LHCb,  BESIII,  EIC, COMPASS, FAIR/GSI, J-PARC, JLAB, NICA,  HERMES, SLAC.
Asymmetry in charged pion production is an important observable in intermediary 
and high  energies 
processes in heavy ions collisions, for example in \cite{partprod1,partprod2}.
Several different physical effects from finite density environment  
may contribute to
 the ratio $\pi^-/\pi^+$ close to three   \cite{Zhang-Rho,JunXu-etal,CMW-pions}
being that it decreases with the beam energy (with $A$ GeV),
and it  increases with  the $N/Z$ asymmetry
\cite{FOPI}.
This  ratio
is related to the dynamics and behavior of hadrons
under different  conditions and it 
helps to determine freeze-out curves 
\cite{Floris}.
There is also an intense and wide investigation in sources of CP breaking at high energies
by analyzing weak interactions  in  heavy meson decays, few examples in
\cite{LHCb-D0D0b,LHCB-D0,babarD0,CDFD0,etapi}.
In the present work, the possibility of CP breaking   in  decays of light axial meson 
is envisaged by means of interference effects.

  Three meson 
interactions based in 
QCD-  one loop  quark polarization  have been calculated  
by considering, for example, the Global Color Model (GCM) or the Nambu-Jona-Lasinio (NJL)-model
\cite{GCGCraig,ERV,PRDPLB,PRD-2022b,FLB-2024b}.
Dynamical Chiral Symmetry Breaking (DChSB) endows constituent quarks with
large quark masses and this makes possible 
large quark mass  expansions of the quark determinant.
Flavor-Dirac quark currents provide suitable states to introduce 
meson multiplets, already  coping with the constituent quark model,
what can be implemented 
by means of the auxiliary field method.
However, 
there are  longstanding 
controversial   results concerning  light  scalar and axial mesons   structures, for which 
non-quark-antiquark states may contribute considerably
\cite{pelaez,PDG,lightAxial}.
Nevertheless, along this work, it will be assumed  that an axial meson nonet 
and a light scalar meson nonet may
be defined, although the physical states may be a result of superposition with  different 
non-quark-antiquark components.
In particular, the following scalar and axial mesons will be considered 
as part of  flavor U(3) nonets:
the isotriplets  scalar $a_0(980)$ and axial $a_1(1260)$, 
and isoscalar axial mesons $f_1(1285)$ and $f_{1S}(1420)$
\cite{PDG}.
The production of axial $f_1(1285)$ and $f_{1S}(1420)$
has been investigated 
in proton-proton collisions 
%in the WA102 Collaboration  
\cite{WA102,Lebiedowicz3,Lebiedowicz2} and it may involve different
types of processes.
In the present work, the decay of  these two axial mesons 
into charged pions and another meson - either a charged $\rho(770)$ or 
charged scalar $a_0(980)$,
will be investigated with three-meson vertices.

Therefore,
in this work, some specific  three light meson leading  interactions 
are re-derived, along the lines  of  \cite{FLB-2024b}
and used to investigate neutral axial meson decays into charged pions.
These flavor U(3) interactions describe several channels, among which 
only  processes involving mesons
composed exclusively by up and down quarks in the  final states will be considered.
The channels with strange mesons will necessarily involve
states for which there are  still further problems to 
define quark-antiquark mesons,  
possibly in possible  scalar and axial meson nonets,
and they will be investigated in  a separate work.
Furthermore, the point-like limit of the three-meson  interactions, 
and the corresponding zero momentum  limit of the form factors,
 will be considered.
This will provide an upper bound for the 
resulting effects. 
The following decays of neutral axial mesons
will be considered 
($A= f_1(1285)$ and $f_{1S}(1420)$): 
$A \to \pi^\mp + a_0^\pm$ 
(processes containing one axial, one scalar and one pseudoscalar mesons will be labeled by ASP)
and 
  $A \to \pi^\mp + \rho^\pm$ (processes containing one axial, one vector and one pseudoscalar mesons will be labeled by VPA). 
To complement the description, the decay $\rho^\pm \to \pi^\pm + \pi^0$
- with large relative branching ratio - will also be
presented.
The work is organized as follows.
In the next section, the  flavor U(3)  quark determinant
with background (constituent) quark currents 
  is considered in the limit of local quark currents
that give rise to   the corresponding light meson 
multiplets  by means of the 
auxiliary field method (AFM).
All the local couplings are resolved by means of a zeroth order derivative expansion
 in the limit of zero momentum exchange, 
leading to coupling constants along the lines presented in \cite{FLB-2024b}.
In section \eqref{sec:VPAASP} three-meson interactions for 
 two channels of neutral axial meson decays,
$f_1(1285)$  and $f_{1S}(1420)$,
 into
charged pions and rhos $\rho^\pm(770)$   
 or  into charged pions and scalars $a_0^\pm(980)$
will be presented, named respectively as VPA and ASP.
For the channel VPA a next leading interaction will be also considered as 
shown in section \eqref{sec:ddVPA}.
In section \eqref{sec:VPP} the  leading interactions describing the decay 
$\rho^\pm \to \pi^\pm \pi^0$,
VPP, 
are  presented.
In sections \eqref{sec:renormalization} and \eqref{sec:mesonstates}
the renormalization conditions fixing free parameters of the AFM and the definition 
of meson states, considered along the work, are presented.
In section \eqref{sec:part-prod}
the different  ratios  $\Gamma_{\pi^+}/\Gamma_{\pi^-}$ for each of    
the  decays are calculated by  including meson mixings.
Numerical results are exhibited  in \eqref{sec:numerics}.
In the last section there is a summary.

\section{ Quark determinant with background quark currents: 
and meson local interactions  }

The quark determinant in the presence of  background constituent quark 
currents  can be written as \cite{PRD-2022b,EPJA-2016,FLB-2024b}:
\begin{eqnarray} \label{Seff-det}  
S_{eff}   &=& - i  \; Tr  \; \ln \; \left\{
 i \left[ {S_0}^{-1}   +
\sum_\phi  a_\phi  J_\phi    \right]
 \right\} ,
\end{eqnarray}
where 
$Tr$ stands for traces of all discrete internal indices 
and integration of  space-time coordinates,
$S_{0}$ is the quark propagator 
 with a 
constituent quark mass due to the DChSB ($M$).
The leading background dressed quark currents
  selected to contribute are the following:
\begin{eqnarray} \label{Rq-j}
\sum_\phi  a_\phi  J_\phi   
=
K_0 \left[  2 R (x-y) 
 ( J_S^i (x,y)
+ 
 i  \gamma_5 \lambda_i  J_{PS}^i (x,y) )
-
  {R}^{\mu\nu} (x-y) \gamma_\mu 
 \lambda_i  ( 
  \gamma_5   J_\mu^{V,i} (x,y)
+ 
 \gamma_5  J_{\mu}^{A,i} (x,y) )
\right],
\end{eqnarray}
where  $\lambda_i$ are the flavor Gell Mann matrices,
 flavor indices of the adjoint representation are  $i,j = 0...(N_f-1)^2$,
 $K_0 = 2 \alpha_g^2/9$ being
$2/9$ from a Fierz transformation and $\alpha_g$ the running
quark-gluon coupling constant.
The   quark currents were defined respectively as
$ J_S^i  = J_S^i (x,y) =  (\bpsi \lambda^i \psi)$,
 $J_{PS}^i = J_{PS}^i (x,y) = (\bpsi i \gamma_5 \lambda^i \psi)$,
$J_{V,i}^\mu = J_{V,i}^\mu (x,y) =  (\bpsi  \gamma^\mu \lambda^i \psi)$ and
 $J_{A,i}^\mu  = J_{A,i}^\mu (x,y) = (\bpsi \gamma^\mu \gamma_5 \lambda^i \psi)$.
Note that, in Eq. \eqref{Seff-det}, the background quark currents are dressed by components of the 
effective gluon propagator  
$R(x-y)$ and $R^{\mu\nu}(x-y)$ that are defined below, and they can be considered as constituent quark currents.
The large quark mass expansion  done below is also compatible with 
a large gluon effective mass expansion since  all the coefficients $a_\phi$ in Eq. \eqref{Seff-det}
contain components of the gluon propagator making the overall quantities $a_\phi J_\phi$ with 
relative small strength.
Although the resulting model is  non-local, the large quark mass
(and to some extent the large effective gluon mass) regime can be considered 
suitable to take
the local limit of such quark currents.
In this work, the indices of the currents 
will be treated such as it does not matter whether
they stand as superscripts or subscripts.
The functions $R(x-y)$ and $R^{\mu\nu}(x-y)$  
are expressed in terms of
 the  longitudinal and transverse components of the gluon propagator 
and they can be written  in momentum space as:  
\begin{eqnarray} \label{gluonpro}
R(k) &=& 3 R_T(k) + R_L(k),
\nonumber
\\
R^{\mu\nu}(k) &=& g^{\mu\nu} (R_T (k) + R_L (k) )
+ 2 \frac{k^\mu k^\nu}{k^2 } (R_T(k) - R_L(k) ).
\end{eqnarray}
There are several ways to treat this determinant.
The quark propagator in Eq. \eqref{Seff-det}  is given by
$S_0(k)  = ( \slashed{k} - M  + i \epsilon )^{-1}$
for which the quark mass is a (large) constituent quark mass, usually considered to be 
generated by DChSB,  but that may
take into account other mechanisms of mass generation.
A  direct large quark (and gluon) effective mass(es) expansion
will be performed
by assuming it has  somewhat large  relative strength with respect to the 
quark fields / currents.
The   dimensionless quantity, that is considered to be small,  is of the following type:
$
S(k) R(k) J_\phi$,
where $S(k)$ is the quark propagator, $R(k)$ a component of the gluon propagator
(that may be parameterized in terms of an effective gluon mass)
and $J_\phi$ is any of the constituent quark  current.
For external lines with larger momenta $P$, as discussed below for the third
order terms of the expansion, the validity of this expansion 
should be  still better
since the contribution of the large momenta in internal lines will
make the quark propagator strength $S_0(k+P)$ still smaller.
Accordingly, although the strength 
of these resulting coupling constants/form factors
will decrease with large momenta, the overall contribution of the 
interactions is more complicated to be analyzed
 because those interactions addressed below are 
explicitely momentum dependent.
Note that this large quark mass expansion allows
the  direct identification of the structure of the resulting interaction terms.
%Effectively, as shown in Refs. \cite{}, the resulting meson-constituent quark coupling constants
%are of the order of magnitude of the phenomenological meson-nucleon coupling constants by considering a single renormalization condition.
For  corresponding calculations with the Nambu-Jona-Lasinio model, for 
example in Refs.
\cite{NJL-U5,PRDPLB},
it has been shown that this expansion provides corrections for 
the NJL-coupling constant of the order of 
$10\% - 25\%$ that is inside the perturbative regime (in the sense of 
large quark mass and strong coupling constant).
A more complete calculation of the determinant, and the full momentum dependence
of the resulting interactions, will be
presented elsewhere.

The auxiliary field method is extremely useful to implement 
quark-antiquark states from the quark currents.
It produces quark model meson multiplets naturally based in flavor symmetry.
However, instead of introducing the non-local auxiliary (meson) fields,
the local limit of the above quark currents will be considered and 
 local meson fields for each of the 
 channels (scalar, pseudoscalar, vector and axial)
 will be introduced
by means of
functional delta functions \cite{AFM-Z}
that can be written as
\begin{eqnarray} \label{AFMdelta}
1 &=&
N \int D [S_i, P_i ]  \delta (
  S_i  -  G_{0,s}^{ii}  J_i )   \delta (  P_i - G_{0,p}^{ii}  J_{P,i} )
\; 
\int D [V_\mu, A_\mu]  \delta  (  V^\mu_i  -  G_{0,v}^{ii}  J_{V,i}^\mu ) 
 \delta  (  A_i^\mu - G_{0,a}^{ii}  J_{A,i}^\mu )
,
\end{eqnarray}
where $N$ is a normalization,
$S_i, P_i, V_i^\mu, A_i^\mu$ are the  meson fields of each of the flavor multiplet,
$D [S_i, P_i ]$ and $D [V_\mu, A_\mu]$
are the measures of integration,
and $G_{0,s}^{ii}, G_{0,p}^{ii}, G_{0,v}^{ii}$ and $G_{0,a}^{ii}$  are  
constant parameters  (dimension $M^{-2}$).
These parameters contain the meson renormalization constants
that are usually introduced with functional delta function such as the following:
$\delta (\phi \sqrt{Z}_\phi - G_0^{ii} J_{i}^\phi )$,
where $\phi$ is a  local meson fields and $G_0$ is a dimensionful constant
  \cite{JPG-2020b,renormaliz-AFM}.
By adopting this latter form, the former expression \eqref{AFMdelta} is recovered
with  redefinition  of the  normalization constant and of $G_0^{ii}$.

The third order large quark mass expansion of the above determinant produces 
 three-meson interactions.
The leading    possible channels  of interaction
 have been presented,  for  the local limit, in \cite{FLB-2024b}
by considering only the non-derivative or  lowest order derivative terms.
In the present work,
four types of leading order (in derivatives)
 vertices containing pseudoscalar field mesons will be selected
with an additional non-leading (NL) vertex for one of these leading ones.
The coupling constants will be resolved within the zero order 
derivative expansion as presented in \cite{mosel}.

\subsection{VPA and ASP vertices: neutral axial mesons decays}
\label{sec:VPAASP}

There are two types of leading vertices that yield an axial meson decay into a final state 
with only up and down quark mesons.
The first has no spatial derivative acting on the fields and contains 
axial-vector-pseudoscalar (VPA) local meson fields.
The second one has one spatial derivative acting in one of the fields and 
contains axial-scalar-pseudoscalar (ASP) mesons.
In the local limit the following interactions are obtained:
\begin{eqnarray} 
\label{6,sb-VPA-mes}
{\cal L}_{6,sb,VPA}^{mes} &=&
 T^{ijk}  
  \frac{ 
3 i \; G_{sb1} \left(
V^{\mu}_{i} A_{\mu}^{j}  P_{k}  
-
 A^{\mu}_{i} V_{\mu}^{j}
 P_{k}
\right)  }{ 
 G_{0,v} G_{0,ps} G_{0,a}
}  
,
\end{eqnarray}
\begin{eqnarray} 
\label{6d-ASP-mes} 
{\cal L}_{6d,ASP}^{mes}
&=&
 3  \;   \frac{  T^{ijk}}{ G_{0,a} G_{0,s} G_{0,ps} }  
\left\{ \left(  \partial_\mu P_i \right) 
\left[ - G_{d1}
 S_{j}  
A^{\mu}_{k}
+ G_{d2} 
A^{\mu}_{j}
 S_{k}  
\right]
- G_{d2} (\partial_\mu S_i)
\left[
 A_j^\mu P_k 
+  P_j  A^\mu_k 
\right] 
\right.
\nonumber
\\
&+&
\left. 
(\partial_\mu A^{\mu}_{i}  )
\left[ 
G_{d1} S_j P_k  + G_{d2} P_j S_k
\right]
\right\},
\end{eqnarray}
where
$T^{ijk} =
 2 (d_{ijk} + i f_{ijk})$, 
 in terms of the flavor group
 symmetric and anti-symmetric
structure constants respectively.
By defining the function $R^\eta (k)$ to be either the longitudinal $\eta=L$ or the 
transverse $\eta=T$ part of the gluon propagator,
the following flavor dependent coupling constants 
 were defined,  for  the limit of 
zero    momentum exchange:
\begin{eqnarray}
\label{G6s}
\label{G6SVV} 
G_{sb1} T_{ijk} \; g^{\mu \rho} 
&=&  
\; A_{2\eta} \frac{2 N_c}{3}  K_0^3  Tr_{D,F}
\int_k 
 {S}_0  (k) \gamma^\mu \lambda_i 
 R^\eta (-k)
{S}_0  (k)
\gamma^\rho \lambda_j  R^\eta  (-k)
 {S}_0  (k)
   \lambda_k
 R^\eta  (k) ,
\nonumber
\\
\label{dSSV}
G_{d1} T_{ijk} g^{\mu\rho}    &=&  
 A_{1\eta} \frac{4 N_c}{3}  K_0^3  Tr_{D,F}
\int_k 
 \tilde{S}_0  (k)  \gamma^\rho  \gamma^\mu \gamma_5 \lambda_i  
 R^\eta  (-k)
 {S}_0  (k) \lambda_j  R^\eta  (-k)
 {S}_0  (k) \lambda_k  \gamma_5 {R}^\eta  (k) ,
\nonumber
\\
\label{dPSVAa}
G_{d2} T_{ijk} g^{\mu\rho}   &=&  
 A_{1\eta} \frac{4 N_c}{3}  K_0^3  Tr_{D,F}
\int_k 
 \tilde{S}_0  (k)  \gamma^\rho  \gamma^\mu \gamma_5 \lambda_i  
 R^\eta (-k)
 {S}_0  (k)  \gamma_5 \lambda_j  R^\eta  (-k)
 {S}_0  (k) \lambda_k   {R}^\eta (k) ,
,
\end{eqnarray}
where  $Tr_{D,F}$ stands for the traces in Dirac and Flavor indices,
for a constituent effective quark mass $M$ one used
$\tilde{S}_0 (k)  = 1/(k^2 - M^2 + i\epsilon)$.
 $S_0(k) = (\slashed{k} - M + i\epsilon)^{-1}$ is the free quark propagator,
and the following coefficients related to the gluon propagator component were defined:
\begin{eqnarray}    \label{ALT}
A_{1T} =\frac{3}{2}, \;\;\;\;
A_{1L} = \frac{1}{2},
\;\;\;
A_{2T} = 3, \;\;\;
A_{2L} = 1. \;\;\;
\end{eqnarray}
It is interesting to note that, in general,
the coupling constants for a particular two meson decay channel are the same for both
types of interactions with symmetric and antisymmetric 
flavor structure constants
$d_{ijk}$ and $f_{ijk}$ .
This means that the specific  momentum dependence 
of such  form factors,  defined by Eqs. \eqref{G6s}
and in other equations below,
 should nearly factorize in the ratios
of decay rates explored below.
Therefore, these momentum dependencies 
should not introduce considerable changes in the resulting ratios
of meson decay,
and the  coupling constants will be considered for numerical 
calculations  instead of the form factors.
Furthermore, this issue must be related to the
problem of fixing the vector and axial meson polarizations for all the corresponding
momentum dependent vertices. 
The inclusion of the full momentum dependent form factors
and their behavior, conjugated to the choice of the vector/axial meson 
polarization, as discussed after Eq. \eqref{A0-A8} and below, 
will be developed in another work.
  After having defined the coupling constants at the zero momentum exchange, i.e.
neglecting the momentum dependence, all the energy dependencies of the couplings in 
Eq. \eqref{6d-ASP-mes} will also be neglected by fixing the longitudinal vector and axial meson polarizations,
such as $\mu=0$, and fixing the meson energy as its rest energy, i.e. $E_\phi = M_\phi$.

All the coupling constants are ultraviolet finite.
Although the resulting effective model is non-renormazible,
effects of these  coupling constants  of meson dynamics make sense by 
means of a suitable renormalization to fix the parameters
$G_{0}^{ii}$,  what is discussed below.
The coupling constant $G_{sb1}$
is  proportional to the quark mass that has been taken as constant.
Therefore, it corresponds to
(chiral) symmetry breaking term, and it disappears in the limit of massless quarks.
$G_{d1},G_{d2}$
are non-zero, even in the case of massless quarks.
Whereas  in the ASP channel   there are  terms of  both types with $f_{ijk}$ and $d_{ijk}$,
 for the channel VPA   only the antisymmetric structure constant
contributes. 
Therefore, a next-leading (NLO) VPA term will  also  be taken into account.

\subsubsection{ NLO VPA }
\label{sec:ddVPA}

One next leading higher order derivative term will be considered now
  that will be referred to as ddVPA.
There are several contributions that, in the local limit,
 can be written as:
\begin{eqnarray} \label{ddVPA} 
{\cal L}_{ddVPA} &=& 
3 \; i \; \frac{T_{ijk}  G_{sva,sb} }{G_{0,p} G_{0,v} G_{0,a} }
\left\{ 
  \Gamma^{\nu\alpha\beta\mu} A_\nu^i (\partial_\alpha P_j) (\partial_\beta V_\mu^k)
+   
\Gamma^{\nu\beta\mu\alpha} A_\nu^i   (\partial_\beta V_\mu^j) (\partial_\alpha P_k)
-  
\Gamma^{\alpha\nu\beta\mu} (\partial_\alpha A_\nu^i )
 P_j    (\partial_\beta   V_\mu^k)
\right.
\nonumber
\\
&-&
\left.  
\Gamma^{\alpha\nu\beta\mu} (\partial_\alpha A_\nu^i )
  (\partial_\beta   V_\mu^j)  P_k
+ \Gamma^{\beta\nu\alpha\mu} (\partial_\beta  A_\nu^i)  (\partial_\alpha P_j) V_\mu^k
- 
\Gamma^{\alpha\beta\nu\mu}  (\partial_\alpha P_i) (\partial_\beta  A_\nu^j)  V_\mu^k
\right\} ,
\end{eqnarray}
where the following tensor was defined:
\begin{eqnarray} \label{tensors}
\Gamma^{\mu\alpha\beta\rho}
&=& 
\left( g^{\mu\alpha}g^{\rho \beta} 
- g^{\alpha\rho} g^{\mu\beta} 
+ g^{\beta\alpha} g^{\mu\rho} \right)
,\end{eqnarray}
and the coupling constant, for the limit of zero momentum exchange, was defined as:
\begin{eqnarray} \label{GSVA}
G_{sva,sb}  T_{ijk}   \Gamma^{\mu\alpha\beta\rho}
&=& 
  \frac{2 N_c}{3} K_0^3 \; Tr_{D,F} \int_k 
 {S}_0   (k) \gamma_{\nu} \gamma^5 \lambda_i R^{\mu\nu} (k) 
\tilde{S}_0   (k) \gamma^{\alpha}  i\gamma_5 \lambda_j R (-k) 
\tilde{S}_0 (k)  \gamma^\alpha i\gamma_\sigma \lambda_k R^{\rho\sigma} (-k) ,
\end{eqnarray}

To deal with such large number of terms,
they were rewritten by performing an integration by parts
to eliminate the derivative $\partial_\alpha V_\mu$.
Besides that, the same remarks written after Eq. \eqref{ALT} applies:
vector and axial meson polarizations were chosen  to be  the longitudinal,
$V_\mu \to V_0$ and $A_\mu \to A_0$.
As a consequence, most of the terms depend directly
on  the pion and axial meson  energies ($E_\pi$ and $E_A$).
However, since the momentum exchange of the couplings was neglected
in their integrals, the energy of the mesons will be fixed as their rest masses.
With these choices,  
the vertices obtained from the above equation can be written as:
\begin{eqnarray} \label{VPA-NLO-dd}
{\cal V}_{ddVPA}
&=& 
- 6 \;  i \;  \frac{ G_{sva,sb} }{G_{0,p} G_{0,v} G_{0,a} }
\left\{ 
2  d_{ijk}  A_0^i V_0^j P^k ( E_\pi E_A +  E_\pi^2 - E_A^2 )
+ 
2   i f _{ijk}  A_0^i V_0^j P^k (  E_\pi  E_A  )
\right\}.
\end{eqnarray}

\subsection{ VPP: Charged rho meson
decay into two pions
 }
\label{sec:VPP}

In this section we develop vertices with one vector meson and two pseudoscalar mesons,
labeled by VPP.
There are several ways to treat the quark determinant.
Strictly,  the expansion shown   above
does not describe several channels, such as the
charged vector meson decays into two pions,
$\rho^\pm \to \pi^\pm \pi^0$.
To do that, 
let us  define the following quantity in the determinant
\begin{eqnarray}
X = S_0 (k) \;  K_0 \;  R^{T,L} (k) \;  J_\phi
\end{eqnarray}
Eq. 
\eqref{Seff-det} will  be worked out, except for an irrelevant constant, as
\begin{eqnarray} \label{detXXdag}
 - i  \; Tr  \; \ln \; \left\{ 
 i [ 1 + X ]  \right\} = 
- \frac{i }{2} Tr \; \ln \; \left\{  [ 1 + X ] [ 1 + X^* ] \right\} .
\end{eqnarray}
The local auxiliary fields can then be introduced in the same way
by reducing the currents to their local limit after resolving the coupling constants.
In the second order expansion, mass terms of charge 
 (and mass) eigenstates as
combinations of the   form:
\begin{eqnarray}
{\cal L} &=& 
- \frac{1}{2} \sum_{\phi}   M_{\phi_i}^2 \phi_i^2 \cdot {\phi_i^*},
\;\;\;\;\;
\phi_i = 
S_i, P_i, V_{\mu,i}, A_{\mu, i} ,
\end{eqnarray}
where $M_{\phi_i}^2$ can be calculated straightforwardly, but 
will not be developed in the present work.
To obtain interactions  that describe explicitly the desired decay,
 three  third order terms with a single pseudoscalar complex conjugate field
will be selected with the single derivative acting in one of them
such as to compose a positive parity Lorentz scalar 
interaction term.
The  terms with such structure, in the local limit,
  can be written as:
\begin{eqnarray} \label{L-VPP-1}
{\cal L}_{VPP}^{mes}  &=&
3 T^{ijk} \left[
   i  G_{d2} \left(   
\left\{  (\partial^\mu V_\mu^i )  P_j^*   P_k \right\} 
-  
\left\{ V_\mu^i  (\partial^\mu P_j^*)  P_k
\right) \right\}
+ 
  i  G_{d1}
\left\{  V_\mu^i  P_j^*   (\partial^\mu  P_k) \right\}
\right].
\end{eqnarray}
In this expression, both types of interactions, with  the symmetric and with the antisymmetric
structure constants, will contribute to the decay $\rho^\pm \to \pi^\pm \pi^0$.

\subsection{ Axial meson mixings and  renormalization conditions}
\label{sec:renormalization}

The parameters $G_{0\phi}$ introduced with the 
local auxiliary meson fields, Eq. \eqref{AFMdelta}, can be fixed by a
suitable renormalization of the meson fields, 
along the lines presented in \cite{JPG-2020b,PRD-2022b,FLB-2024b}.
For that, consider the following second order 
meson kinetic  terms obtained from the determinant expansion:
\begin{eqnarray}
{\cal L}_{kin} 
&=& 
\frac{1}{2} \frac{  I_{0,S}^{ii}}{ (G_{0,S}^{ii})^2 } 
 \partial_\mu S_i  \partial^\mu S_i 
+
\frac{1}{2} \frac{  I_{0,S}^{ii}}{ (G_{0,P}^{ii})^2 } 
 \partial_\mu P_i  \partial^\mu P_i 
-
\frac{1}{8} \frac{  I_{0,V}^{ii}}{(G_{0,v}^{ii})^2 }
 {\cal F}_i^{\mu\nu} {\cal F}_{\mu\nu}^i  
-
\frac{1}{8} \frac{  I_{0,A}^{ii}}{(G_{0,a}^{ii})^2 }
 {\cal G}_i^{\mu\nu} {\cal G}_{\mu\nu}^i  ,
 \end{eqnarray}
where the Abelian vector and axial meson tensors were defined 
as 
$$
{\cal F}_{\mu\nu}^i = \partial_ \mu V_\nu^i - \partial_\nu V_\mu^i ,
%+ [ V_\mu , V_\nu ]  + [ A_\mu , A_\nu ],
\;\;\;\;\;\;\;
{\cal G}_{\mu\nu} = \partial_ \mu A_\nu^i - \partial_\nu A_\mu^i.
%+ [ V_\mu , A_\nu ],
$$
Some of the neutral  meson fields develop mixing interactions
  totally analogous to the vector or pseudoscalar meson mixings
shown in \cite{JPG-2020b,PRD-2021,JPG-2022}.
The following mixing terms  will be considered:
\begin{eqnarray} \label{axialmixings-1}
{\cal L}_{mix} 
&=& 
 \frac{ I_{i \neq j}^{0,\phi} }{ G_{0,\phi}^{ii} G_{0,\phi}^{jj} }
\phi^i \phi_j
\;\;\;\;\;
i,j = 0, 3, 8, \;\;\;\;\; \phi = S, P, V, A.
\end{eqnarray}
Although
 energy dependent vector and axial meson mixings also arise, i.e.
$\sim {\cal G}_i^{\mu\nu} {\cal G}_{\mu\nu}^j$, 
they are  equivalent to the former by a field redefinition.
In the present work, the energy independent limit of the interactions was adopted,
since the coupling constants were defined in the zero momentum exchange limit.

The above equations  fix the parameters $G_{0,\phi}$
by means of an ultraviolet cutoff adopted 
for the quantities $I_{0,\phi}^{ii}$.
These terms  give rise to the canonical field normalization  terms,
and this procedure settles the parameters $G_0^{ii}$.
Consequently,   renormalized  
three-meson coupling constants, by omitting 
the flavor indices $^{ii}$ that are all equal for the case of 
degenerate quark mass, can be defined, such as:
\begin{eqnarray}
\label{mesonfielddef}
G_{sb1}^{R} =  \frac{  G_{sb1}}{ G_{0,v} G_{0,p} G_{0,a}} ,
\;\;\;\;\;
G_{d1/2}^{R} =  \frac{   G_{d1/2}}{G_{0,s} G_{0,p} G_{0,a}} , 
 \;\;\;\;\;  
G_{sva,sb}^R =   \frac{ G_{sva,sb}  }{  G_{0,v}  G_{0,p}  G_{0,a}}.
\end{eqnarray}

The 
 renormalized mixing  parameters, with dimension $M^2$, can be written as:
\begin{eqnarray}
G^R_{i\neq  j}
&=&  \frac{  I_{0,\phi}^{i j} }{  G_{0,\phi}^{ii} G_{0,\phi}^{jj} } , \;\;\;\;\;
i,j = 0, 3, 8.
\end{eqnarray}
Below, the mixing will be considered for the axial meson sector and, in a particular case,
for the pseudoscalar $\pi^0-\eta$.

\subsection{ Meson states }
\label{sec:mesonstates}

The description of light scalars structure has   controversial results.
However, it will be considered that  the
isotriplet states $S_{1,2,3}$ are defined as quark-antiquark mesons 
$a_0(980)$ \cite{PDG}.
The axial nonet is not well established neither \cite{PDG,lightAxial}
and to avoid further uncertainties in the calculation,  the analysis will be restricted 
to final states containing only 
mesons with up and down quarks. 
Whereas the flavor axial meson  nonet will be denoted by $A_i$ or $A_1^i$ ($i=0, 1,..8$),
the isotriplet axial mesons  will be denoted by $a_1(1260)$.
The following  states will be considered:
\cite{PDG,hiller-etal}:
\begin{eqnarray}
\frac{P_1 \pm i P_2}{\sqrt{2}} \sim   \pi^\pm (140), 
&& 
P_3 \sim \pi^0 (135), 
\nonumber
\\
\frac{V_1 \pm i  V_2}{\sqrt{2}}  \sim   \rho^\pm (770), 
 && 
V_3 \sim \rho^0 (770),
\;\;\;\;\;\; \Gamma = 0.147 \; GeV
\nonumber
\\
\frac{A_1 \pm  i A_2}{\sqrt{2}}  \sim   a_1^\pm (1260), 
 && 
A_3 \sim a_1^0 (1260),
\;\;\;\;\;\; \Gamma = 0.420 \; GeV,
\nonumber
\\
\frac{S_1 \pm  i S_2}{\sqrt{2}}  \sim   a_0^\pm (980), 
 && 
S_0 \sim a_0 (980),
\;\;\;\;\;\; \Gamma = 0.075 \; GeV.
\nonumber
\\
S_0 \sim \eta (548) &&  M_{\eta} =  548 \; GeV , \;\;\;\;\;
\Gamma_\eta = 1.3 \times 10^{-6} \; GeV.
\end{eqnarray}
 Besides that, 
the following states for the  corresponding neutral  axial mesons, with their masses 
and widths,   will be considered below   \cite{PDG}:
\begin{eqnarray}  \label{mesons}
A_0 &\sim& f_1 (1285), \;\;\; M_{f_1} = 1.252 \;GeV , \;\;\;\;
\Gamma = 0.023 \; GeV,
%\nonumber
%\\
%A_3 (A^0_1)
% &\sim&  a_1(1260), \;\; M_{A_1} = 1.230 \;GeV  ,\;\;\; 
%\Gamma = 0.40 \; GeV,
\nonumber
\\
A_8 &\sim&  f_{1S} (1420), \;\; M_{f_{1S}} = 1.426 \;GeV  , \;\;\;
\Gamma = 0.055 \; GeV.
%\nonumber
%\\
%V_0 &\sim& \omega (782), \;\; M_{\omega} =0.782 \;GeV , \;\;\;
%\Gamma = 0.008 \; GeV,
%\nonumber
%\\
%V_3 (\rho^0)
% &\sim&  \rho(770), \;\; M_{\rho} = 0.775 \;GeV  ,\;\;\; 
%\Gamma = 0.149 \; GeV,
%\nonumber
%\\
%V_8 &\sim&  \phi (1020), \;\; M_{\phi} = 1.019 \;GeV  , \;\;\;
%\Gamma = 0.004 \; GeV
%\nonumber
%\\
%S_0 &\sim&  \sigma (500), \;\;\; M_{\sigma} = 0.60 \;GeV  , \;\;\;
%\Gamma = 0.28 \; GeV,
%\nonumber
%\\
%S_3 (a_0^0) &\sim& a_0^0 (980), \;\;\; M_{a_0} =  0.980 \; GeV , \;\;\;
%\Gamma = 0.075 \; GeV,
%\nonumber
%\\
% S_8 &\sim&  f_{0} (980), \;\;\; M_{f_{0}} = 0.99  \;GeV , \;\;\;
%\Gamma = 0.050 \; GeV,
\end{eqnarray}
where the widths were adopted as an average 
value of the range presented in PDG Table.

\section{ Asymmetries in the decay of  light axial mesons into charged pions }
\label{sec:part-prod}

\begin{figure}[ht!]
\centering
\includegraphics[width=110mm]{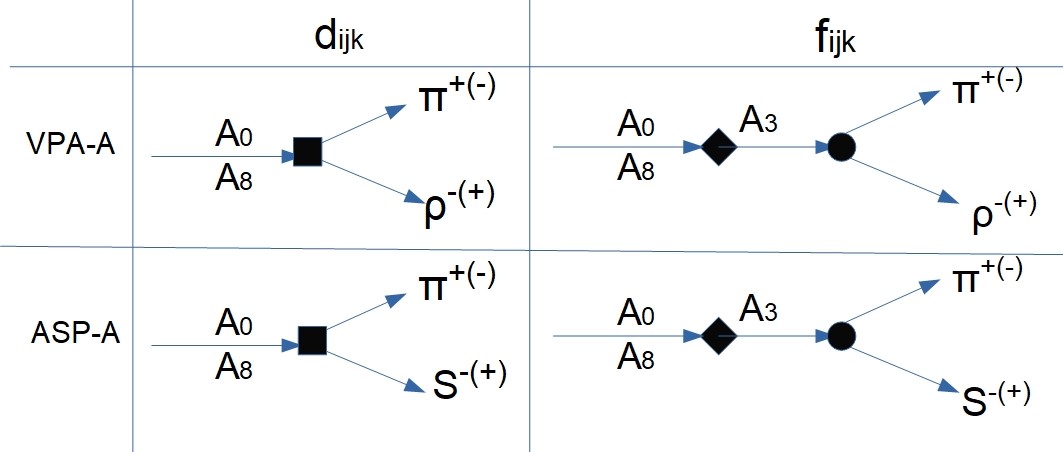}
\caption{ \label{fig:VPAAASP}
\small
Three leg meson decays 
of two neutral  axial  mesons $A_0 \sim f_1(1185)$
and $A_8 \sim f_{1S} (1420)$,
(\ref{mesons}), with   VPA and ddVPA interactions.
 There are d-channel (square) and f-channel (round) couplings involving
charged pions.
Intermediary states from axial meson mixings (diamonds) 
are considered in the antisymmetric channels.
 }
\end{figure}

\begin{figure}[ht!]
\centering
\includegraphics[width=110mm]{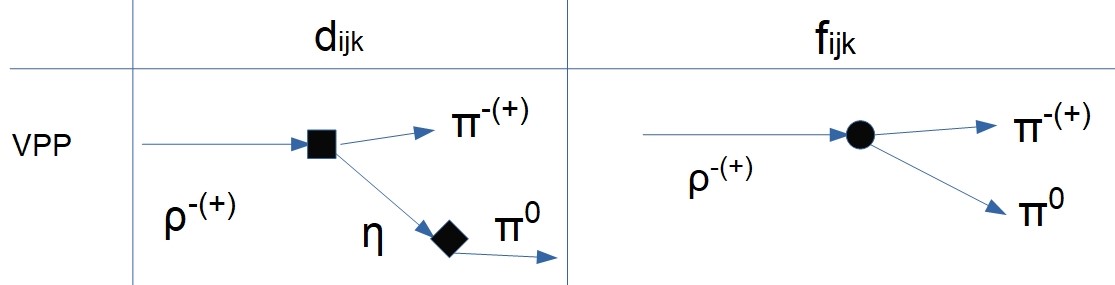}
\caption{ \label{fig:VPPdia}
\small
Three leg meson decays 
of rho into two pions  $\rho^\pm \to \pi^\pm \pi^0$.
The rho decay  is considered with a $\eta-\pi^0$ mixing, with diamond symbol,
in the d-channel.
 }
\end{figure}

\begin{figure}[ht!]
\centering
\includegraphics[width=110mm]{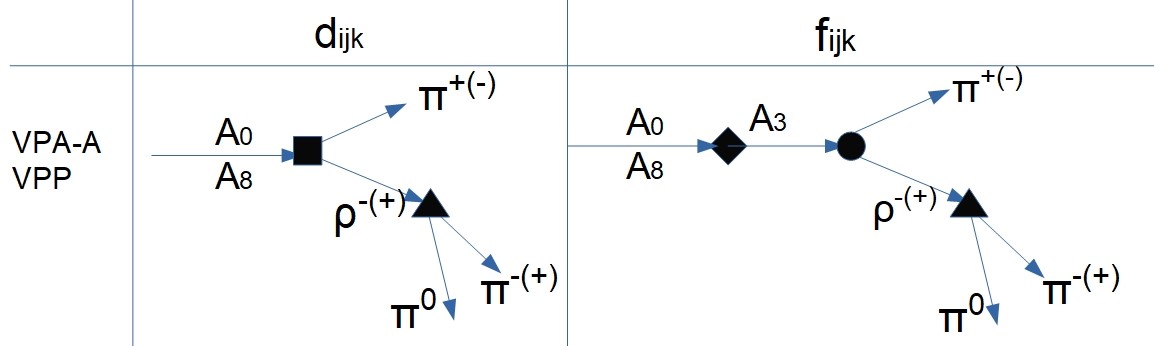}
\caption{ \label{fig:VPAsequencial}
\small
Three leg meson decays 
of two neutral  axial  mesons $A_0 \sim f_1(1185)$
and $A_8 \sim f_{1S} (1420)$ with   VPA and ddVPA interactions.
The d-channel (square) - lhs -  and
f-channel (round) - rhs -  couplings involving
charged pions.
The charged rho is assumed to decay either into $\eta-\pi$ (d-channel)  - 
is encoded in 
triangle symbol, and it includes both VPP d- and f- channels - shown in 
Fig. \eqref{fig:VPPdia}.
 }
\end{figure}

In Figs. \eqref{fig:VPAAASP}, \eqref{fig:VPPdia} and \eqref{fig:VPAsequencial} 
 the specific processes that will be detailed and analyzed below 
are drawn: for the channels VPA, ASP and VPP.
Figs. \eqref{fig:VPAAASP} and \eqref{fig:VPAsequencial}
show the decays of the two axial mesons $A_0$ and $A_8$
 in different channels. 
 The last one, Fig. \eqref{fig:VPAsequencial}, 
has the secondary decay of the rho into two pions,
that is presented separately in Fig.   \eqref{fig:VPPdia}.
In the left-hand side (l.h.s.) columns of all these figures,
there are diagrams obtained for 
 the symmetric structure constant (d-channel) with  square vertices.
They correspond to two different   interactions  (ddVPA and ASP) 
of the 
neutral axial mesons $A_0 \sim f_1 (1285)$ and $A_8 \sim f_{1S} (1420)$.
In the right-hand side (l.h.s.) the amplitudes with the interactions
defined with the antisymmetric structure constant (f-channel) with  round vertices.
Since the decays of axial mesons will be picked up among all the possible channels of the 
interactions VPA and ASP, the corresponding processes may be referred as
VPA-A and ASP-A.
The initial states considered below have   heavier particles than the final state,
and   only  energetically favorable decays will be considered.
In some  cases, due to the flavor channels involved,
a meson mixing (diamond) is assumed to take place such that
both d- and f-channels  yield the same initial and  final states.
In Fig.  \eqref{fig:VPAsequencial}  the  rho meson decay  vertices
are  indicated with a triangle that contains two components,   d-channel and  f-channel 
amplitude.

\FloatBarrier

\FloatBarrier

\FloatBarrier

Most  of the axial meson decays considered in this work
 have already been measured experimentally.
The only decays for which no record was found in the PDG Table
\cite{PDG}
is 
$a_1(1260) \to a_0^0 (980) \pi$ and $f_{1S}(1420) \to \rho \pi$.
The other processes,  showed in figures  \eqref{fig:VPAAASP}, \eqref{fig:VPPdia} and 
\eqref{fig:VPAsequencial} are the following
\cite{PDG}:
\begin{eqnarray} \label{decays-exp}
  &&  a_1(1260) \to \rho (770) \pi,  \;\; \cite{A1rhopi}
\;\;\;\; 
 f_{1S} (1420)  \to a_0 (980) \pi \;\; \cite{WA102,f1sa0pi},
\\
   && 
%f_{1S} (1420) \sim A_8  \to \rho \pi, \cite{f1srhopi}
%\;\;\;\;
f_1(1285)  \sim A_0 \to \rho \pi,  \;\; \cite{f1rhopi}
\;\;\;\;\; f_1(1285)   \to a_0 (980) \pi, \;\; \cite{WA102,BES-Belle,f1a0pi}
\nonumber
\end{eqnarray}
 Some of these decay may respect the OZI sum rule or G-parity 
conservation
and others do not \cite{lightAxial}.
%The decays considered  for the $f_{1S}(1420)$ meson 
%have not been observed experimentally and this might be due to several reasons such
%as very small branching ratio, difficulties with final state interactions 
%or even that its structure does not correspond exactly to the state $A_8$ of a
%quark-antiquark nonet.
%However, 
%it might also happen that 
% state  $A_8$ of the possible axial meson nonet can undergo strong mixings
%with the states $A_0$ and $A_3$ as discussed below 
%and these 
Besides that, neutral axial meson mixings might 
 prevent a clear identification of their structure at the 
example of the $\eta-\eta'$ mixing of the pseudoscalar mesons multiplet
\cite{etaetaprime1,etaetaprime2}.
Therefore, the channels presented above will
 be investigated below, at least,  as a possible scenario.
Besides that, there are decays of the charged rho
$\rho$:
\begin{eqnarray} \label{decays-exp2}
&& \;\;\;\;\rho^\pm  \to \pi^\pm \pi^0, \;\;\;\; (\Gamma_i/\Gamma \sim 100 \%),
\;\;\;\;\;\;\;\;\;\;\; 
\;
\rho^\pm \to \pi^\pm \eta, \;\;\;\; (\Gamma_i/\Gamma \sim 6 \times 10^{-3} \%).
\nonumber
\end{eqnarray}

\subsection{ VPA-A decay}

The  following decays will be analyzed in this section:
\begin{eqnarray}
\label{A0-A8}
A_1^8 &\to& \rho^+ \pi^- , \;\;\;\;\; A_1^8 \to \rho^- \pi^+ ,
\nonumber
\\
A_1^0 &\to& \rho^+ \pi^- , \;\;\;\;\; A_1^0 \to \rho^- \pi^+ .
\end{eqnarray}

Consider the antisymmetric part of the  $VPA$ coupling, in  Eq. (\ref{6,sb-VPA-mes}),
and  of the ddVPA contribution, in Eq. \eqref{VPA-NLO-dd}, with 
 $T_{ijk} \to  2 i f_{ijk}$.
The vector and axial mesons will be considered in  their longitudinal polarizations, $\mu=0$,
such that they will be omitted.
As discussed above, the energy dependence of the couplings will be reduced to the
meson rest energies.
By adding both contributions for the f-channel, it yields:
\begin{eqnarray} \label{VPA-f1}
{\cal L}_{VPA}^f = 
6 \; i  \; f_{ijk} 
\left[
2  G_{sb1}^R
+  2 E_\pi E_A   \;  G_{sva,sb}^R
\right] V_{i} A^{j}  P_{k}  
,
\end{eqnarray}
These terms
 can be written in terms of the corresponding
mesons as charge/mass eigenstates with 
 up and down quarks, i.e. for $i,j,k= 1,2,3$.
It yields:
\begin{eqnarray}   \label{VPA-f2}
\label{VPA-asymm}
{\cal L}_{A \rho^\mp \pi^\pm}^{f}
=
 12  \;  i \; 
\left[
 G_{sb1}^R
+   E_\pi E_A   \;  G_{sva,sb}^R
\right]
\left[ 
 \pi_3 ( \rho^- a_1^+ - \rho^+ a_1^- )
+   a_1^0 ( \rho^+ \pi^-  - \rho^- \pi^+ )
+ \rho_3 ( a_1^- \pi^+ - a_1^+ \pi^- )  
\right],
\end{eqnarray}
where $\pi, \rho$ and $a_1$ are 
  components  of the isotriplets: respectively pion, rho and axial $a_1$.
The second interaction in this last equation
corresponds to the r.h.s. of   VPA-A in Fig. \eqref{fig:VPAAASP}
before introducing the axial meson mixing that will be discussed below.
The decay widths of the $f_1(1285)$ and of the $f_{1S}(1420)$ are not
very large so that the mixing may be expected to occur.
Note that there are different signs in the contribution of 
each of the positive and negative mesons for each coupling.

A related amplitude     appears from the contributions of 
the   vertex  NLO-VPA (ddVPA) with 
the symmetric structure tensor, $d_{ijk}$.
These terms 
 can be written as: 
\begin{eqnarray} \label{VPA-symm}
{\cal L}_{VPA}^{d}
&=&   
- 12  \;  i \;  d_{ijk} \;  G_{sva,sb}^R  (   E_A E_\pi  + E_\pi^2 - E_A^2) \; 
  A^i V^j P^k 
.
\end{eqnarray}
for which  the following terms will be considered
$d_{118} = d_{228} = \frac{1}{\sqrt{3}}$
and $d_{110} = d_{220}  = \sqrt{\frac{2}{3}}$
respectively 
for axial mesons $A_0$ and $A_8$ decaying into
 charged pions.
The  processes of the 
l.h.s. of the first line of Fig. \eqref{fig:VPAAASP},
correspond therefore to the following interactions:
\begin{eqnarray} \label{TsymVPA}
{\cal L}_{A \rho^\mp \pi^\pm}^{d}
&=&
- 12  \;  i \;   G_{sva,sb}^R  (   E_A E_\pi  + E_\pi^2 - E_A^2)
\left[ 
\frac{A_8}{\sqrt{3} }
\left( 
\rho^+ \pi^- + \rho^- \pi^+
\right)
+   \sqrt{\frac{2}{3}} A_0 
\left( 
\rho^+ \pi^- + \rho^- \pi^+
\right)
\right]
.
\end{eqnarray}

\subsubsection{ Meson mixing}
\label{sec:mixing}

Consider the  mixings between  neutral meson flavor-eigenstates,
obtained with meson auxiliary fields in the corresponding quark-antiquark interactions
of the following form:
\begin{eqnarray} \label{Gij}
{\cal L}_{mix} =  \sum_{ij}  G_{i j}^R \phi_i \phi_j, \;\;\;\;\;\;\;\;\;
G_{i\neq j} \neq 0 \;\; \mbox{for} \;\;i,j = 0, 3, 8,
\end{eqnarray}
The  pseudoscalar and scalar mixing interactions $G_{i\neq j}$ have been calculated 
and used in 
\cite{PRD-2021,JPG-2022,JPG-2023}.
 These mixings give rise to  
 to the neutral meson  physical mass eigenstates $\varphi_i$ by means of a 
flavor rotation parameterized in terms of $\bar{G}_{ì\neq j}$ such that:
\begin{eqnarray}
\label{meson-mixings}
\phi_0 &\sim& \varphi_0 + \bar{G}_{03} \varphi_3 + 
\bar{G}_{08} \varphi_8 ,
\nonumber
\\
\phi_3 &\sim& \varphi_3 + \bar{G}_{30} \varphi_0 + 
\bar{G}_{38} \varphi_8 ,
\nonumber
\\
\phi_8 &\sim& \varphi_8 + \bar{G}_{80} \varphi_0 + 
\bar{G}_{83} \varphi_3 .
\end{eqnarray}
The neutral axial meson states, from $A^{(0,3,8)}$,
can  be approximated as:
\begin{eqnarray}
A_8  
&\sim&  f_{1S} (1420)
+ 
  \bar{G}_{38} a_1^0 (1260) + \bar{G}_{80} f_1 (1285),
\nonumber
\\
A_0
&\sim&  f_1 (1285)
+ 
  \bar{G}_{30} a_1^0 (1260) 
+ \bar{G}_{80} f_{1S} (1420).
\end{eqnarray}

These mixings  can be applied in the 
Lagrangian terms of the form shown above:
\begin{eqnarray}
{\cal L}_{mix} =  \sum_{ij=0,3,8} G_{ij}^R  A_i A_j.
\end{eqnarray}
By requiring the   new mixings among
mass  eigenstates $\varphi_i \varphi_{j\neq i}$
to disappear,
and by picking up only each of the leading terms of each channel,
the following meson mixing parameters   are obtained:
\begin{eqnarray} \label{Gmixbar}
\bar{G}_{03} \simeq \frac{G_{03}^R}{2 ( G_{33}^R + G_{00}^R) } ,
\;\;\;
\bar{G}_{08} \simeq \frac{G_{08}^R}{2 ( G_{88}^R + G_{00}^R)} ,
\;\;\;
\bar{G}_{38} \simeq \frac{G_{38}^R}{2 ( G_{33}^R + G_{88}^R) } .
\end{eqnarray}

The  resulting f-channel  complete VPA-A amplitudes,
 from eq. \eqref{VPA-f2},
for the three leg vertices with neutral axial meson,
with the mixing $A_8 \to A_3$ and for $A_0 \to A_3$,
are  the following:
\begin{eqnarray}
\label{VAP-pm}
{\cal V}_{A\rho^+ \pi^-}^f
&:& 
12  \; i \; \left[   G_{sb1}^R  
+    E_\pi E_A  \;  G_{sva,sb}^R
\right]
\;
D_{A_3} (P^2) \;
\left( 
   A_1^{8}  \; \bar{G}_{38} 
+ 
A^0_1\;
  \bar{G}_{03}
\right)
,
\nonumber
\\ 
{\cal V}_{A\rho^- \pi^+}^f
&:& 
 - 12  \; i \; \left[   G_{sb1}^R  
+    E_\pi E_A  \;  G_{sva,sb}^R
\right]
\;
D_{A_3} (P^2) \;
\left( 
   A_1^{8}  \; \bar{G}_{38} 
+ 
A^0_1 \;
  \bar{G}_{03}
\right)
.
\end{eqnarray}
where $D_{A_3} (P^2)$ is the intermediary state propagator,  $A_3 \sim a_1^0(1260)$, 
 that 
will be written as a Breit Wigner propagator:
\begin{eqnarray}
D_{\phi} (P^2) = 
\frac{i }{ P^2 - M_{\phi}^2 + i \; M_{\phi} \; \Gamma_{\phi} },
\end{eqnarray}
where $\phi \sim A_3 \sim a_1(1260)$,
in terms of its  mass $M_\phi$ and decay width $\Gamma_\phi$.

The     VPA amplitudes  
can be written in terms of 
  the following quantities  for the f- and d-channel:
\begin{eqnarray}
\label{ASP-GfGd}
& G_{A_8}^f (E_A) = 
 12   \; \bar{G}_{38} \; \left(   G_{sb1}^R  
+   E_\pi E_A   \;  G_{sva,sb}^R
\right)
,
\;\;\;\;\;
& G_{A_0}^f  (E_A) =
12  \; \bar{G}_{03}    \; \left(   G_{sb1}^R  
+   E_\pi E_A   \;  G_{sva,sb}^R
\right)
,
\\
& G_{A_8}^d  (E_A) =
 - \frac{12}{\sqrt{3} }   \; 
  G_{sva,sb}^R  (  E_\pi  E_A +  E_\pi^2  - E_A^2 ) \; 
,
\;\;\;\;\;
& G_{A_0}^d (E_A) = 
- 24   \sqrt{\frac{2}{3}}   \; 
  G_{sva,sb}^R (  E_\pi  E_A +  E_\pi^2  - E_A^2 )  \;   
.
\nonumber
\end{eqnarray}

For the corresponding ratio of decay rates
  for the initial  neutral  axial vector 
with  energy $P$ for each axial meson decay $A_J$ (J=0,8),
 and final states with positive or negative pions
and  rhos,  
 one can write:
\begin{eqnarray} \label{VAP-pi+}
 \label{VAPV-pi+} 
\mbox{Ratio} \;  (VPA - A_J)  \;  : \;   
\frac{ \Gamma_{\pi^+}^J  }{ 
\Gamma_{\pi^-}^J 
}
&\sim& 
\frac{ 
\left|
+ G_{A_J}^d 
+ G_{A_J}^f D_{A_3} (P^2) 
\right|^2
}{
\left|
+ G_{A_J}^d 
- G_{A_J}^f D_{A_3} (P^2) 
\right|^2},
\;\;\;\;\;\;\;\;\;\;
J= 0,8.
\end{eqnarray}

\subsection{ ASP-A  decay}

The  following ASP-A decays will be analyzed:
\begin{eqnarray}
A_1^8 &\to& a_0^+ \pi^- , \;\;\;\;\; A_1^8 \to a_0^- \pi^+ ,
\nonumber
\\
A_1^0 &\to& a_0^+ \pi^- , \;\;\;\;\; A_1^0 \to a_0^- \pi^+ .
\end{eqnarray}

By performing an integration by parts  for the term with $\partial_\mu S_k$ in Eq. 
\eqref{6d-ASP-mes}
all the derivative can be taken to act in the fields P or A.
They can be written  as:
\begin{eqnarray} 
\label{6d-SSV1-mes-2}  
{\cal L}_{6d,ASP}^{mes} &=&
 3  \;    T^{ijk} 
\left\{ \left(  \partial_\mu P_i \right) 
\left[ (G_{d2}^R - G_{d1}^R)
 S_{j}  
A^{\mu}_{k}
+ 2 G_{d2} 
A^{\mu}_{j}
 S_{k}  
\right]
\right.
\nonumber
\\
&+&
\left. 
(\partial_\mu A^{\mu}_{i}  )
\left[ 
(G_{d1}^R + G_{d2}^R) S_j P_k  +  2 G_{d2}^R P_j S_k
\right]
\right\}.
\end{eqnarray}
The polarization of the  axial meson 
will be  longitudinal, $\mu = 0$, and it will be omitted from here on.
The term with the 
antisymmetric structure constant, with states containing only up and down quarks,
involves $A_3 \sim A_1^{(0)}$, the third component of the 
isotriplet $a_1(1260)$, similarly to the VPA channel.
 By  selecting the channels involving charged pions, with $f_{123} = 1$,
and the neutral axial meson,
it yields  the following terms:
\begin{eqnarray} 
\label{GASP-pm-fijk} 
{\cal L}_{A_1 \pi^\pm S^\mp}^f
&=&
 6  \; i  \; G_{S^f} \;
A_1^{3} \left[
S^{-}  ( \pi^+ )
-
S^{+}   (  \pi^- ) 
\right]
,
\end{eqnarray} 
where 
\begin{eqnarray}
G_{S^f}  = G_{S^f}  (E_A)&=& \left[ E_\pi (G_{d2}^R + G_{d1}^R )   
- E_A (G_{d1}^R - G_{d2}^R )
\right].
\end{eqnarray}
There are other channels involving  pions, scalar and  axial mesons
that will not be analyzed.
By considering the mixings $A_1^8-A_1^3$ and $A_1^0-A_1^3$
for the neutral axial meson, 
the corresponding amplitudes can be written as:
\begin{eqnarray}
\label{GASP-pm2} 
{\cal V}_{A  \pi^+ S^-}^f
&:&
6  \; i \;  
G_{S^f} \; D_{A_3} (P^2)
\left(
 \bar{G}_{38}  A_8 +  \bar{G}_{03}  A_0
\right) ,
\nonumber
\\
{\cal V}_{A  \pi^- S^+}^f
&:&
- 6  \; i \;  
G_{S^f} \; D_{A_3} (P^2)
\left(
 \bar{G}_{38}  A_8 +  \bar{G}_{03}  A_0
\right) .
\end{eqnarray}
From this expression   the following quantities are defined:
\begin{eqnarray}
G_{S^f_0} &=& 6  \; G_{S^f}    \bar{G}_{03} ,
\;\;\;\;\;
G_{S^f_8}  = 6 \;  G_{S^f}   \bar{G}_{38} .
\end{eqnarray}

For the symmetric interactions,  
since $d_{11n} = d_{22n}$ (n=0,8),
one has the following amplitudes
with charged pions and scalars and neutral axial meson:
\begin{eqnarray}
\label{GASP-dijk} 
{\cal V}_{A   \pi^\pm S^\mp}^d
&=&
+ \; i \; 
G_{S^d_J} \;
{A_1}^{J}  \;
,  \;\;\;\;\; J = 0, 8,
\end{eqnarray}
where 
\begin{eqnarray}
G_{S^d_J} = G_{S^d_J} (E_A) &=& 
6 \;  d_{11 J}  \; 
\left[ E_\pi ( 3 G_{d2}^R - G_{d1}^R )   
- E_A ( 3 G_{d2}^R + G_{d1}^R )
\right] ,
\;\;\;\;\; J = 0, 8.
\end{eqnarray}

For the corresponding energy dependent decay rates of the  axial vector meson 
 decay into a positive or  a negative pion and scalar,
  the following ratio can be defined:
\begin{eqnarray} \label{ASP-A-pi+}
\mbox{Ratio} (ASP-A_J)  \;  : \; 
\frac{\Gamma_{\pi^+}^J  }{ 
\Gamma_{\pi^-}^J  }
&\sim& 
\frac{
\left|
+ G_{S^d_J} (P^2) + G_{S^f_J} (P^2)  D_{A_3} (P^2)
\right|^2
}{
\left|
+ G_{S^d_J} (P^2)
- G_{S^f_J} (P^2)  D_{A_3} (P^2)
\right|^2},
\;\;\;\;\;\;\;\;\;
 J = 0, 8.
\end{eqnarray}
 
\subsection{
VPP: $\rho^\pm$ decay}

Consider the interaction terms given in \eqref{L-VPP-1}.
By picking up only the f-channel terms describing the decay $\rho^\pm \to \pi^\pm \pi^0$,
it yields:
\begin{eqnarray}
\label{rhopipi} 
{\cal L}_{\rho^\pm \pi^0 \pi^\pm}^f
&=& 
   6  \; i \; f_{123}
\left[
    G_{d2} 
\left(
- (\partial^\mu \rho^+ )
\pi^+ \eta + 
(\partial^\mu \rho^- )
\pi^- \eta 
\right)
-
\left(
-
 \rho^+ 
(\partial^\mu 
\pi^+ ) \eta + 
 \rho^- 
(\partial^\mu \pi^-) \eta 
\right)
\right.
\nonumber
\\
&+& \left. 
G_{d1} \left( 
- \rho^+ 
\pi^+  (\partial^\mu   \eta  )
+
 \rho^- 
\pi^- (\partial^\mu   \eta )
\right)
\right],
\end{eqnarray}
The decay $\rho^\pm \to \pi^\pm \eta$ is described by the following terms:
\begin{eqnarray}
\label{VPP-f3}
{\cal L}_{\rho\pi\pi}^f
&=&
G_V^f
\left[
-  \rho^+ 
\pi^+ \eta +
  \rho^- 
\pi^- \eta 
\right],
\end{eqnarray}
where
\begin{eqnarray}
\label{GVf}
G_{V^f} = 6 
\left[
  ( E_\rho + E_{\pi^\pm} )   G_{d2} 
-  E_{\pi^0}  G_{d1} 
\right].
\end{eqnarray}
As discussed above, 
in this work, the momentum dependence of the coupling constants are not taken into account.
So, accordingly,
 meson energies in these equations are  taken as their rest masses.

Interactions \eqref{L-VPP-1} 
also describe the decay $\rho^\pm \to \pi^\pm \eta$
in the channel with $T^{ijk} \to 2 d^{ijk}$
that, with a mixing $\eta - \pi^0$, leads
 to the same final state 
as the decay \eqref{rhopipi},
\begin{eqnarray}
 \label{rhopieta-1}
{\cal L}_{\rho^\pm \eta \pi^\pm}^d
&=& 
  6 \; i \;  d_{11 0}
\left[
    G_{d2} 
\left(
(\partial^\mu \rho^+ )
\pi^+ \eta + 
(\partial^\mu \rho^- )
\pi^- \eta 
\right)
-
\left( \rho^+ 
(\partial^\mu 
\pi^+ ) \eta + 
 \rho^- 
(\partial^\mu \pi^-) \eta 
\right)
\right.
\nonumber
\\
&+& \left. 
G_{d1} \left( \rho^+ 
\pi^+  (\partial^\mu   \eta + )
+
 \rho^- 
\pi^- (\partial^\mu   \eta )
\right)
\right].
\end{eqnarray}
The mixing $\eta' - \pi^0$ (channel $J=8$) is not considered 
because it is not energetically favorable, i.e. $M_{\eta'} > M_{\rho}$.

By including the meson mixing in the
d-channel,
 the amplitudes can be written as:
\begin{eqnarray}
 \label{rhopieta-1}
{\cal V}_{\rho^\pm \pi^\pm \pi}^d
&=& 
G_{V^d} \;  D_{\eta} (P^2_\eta)
,
\end{eqnarray}
where 
\begin{eqnarray}
\label{GVd}
G_{V^d} =  6 \sqrt{\frac{2}{3}} 
\; \bar{G}_{03} 
 \;
\left[
    G_{d2}   ( E_\rho   + E_\pi )  - E_\eta G_{d1}  \right].
\end{eqnarray}
The $\eta-\pi^0$ mixing in the final state of the $\rho^\pm$ decay can be 
searched experimentally in spite of being a  small effect.
 
The following ratio of decay rates for the charged rho 
can be extracted:
\begin{eqnarray} \label{VPP-ratio}
\mbox{Ratio} (VPP)  \;  : \; 
\frac{\Gamma_{\pi^+}^V  }{ 
\Gamma_{\pi^-}^V  }
&\sim& 
\frac{
\left|
+ G_{V^d}     D_{\eta} (P^2)  - G_{V^f}  
\right|^2
}{
\left|
+ G_{V^d}     D_{\eta} (P^2)
+ G_{V^f}  
\right|^2}
\;\; .
\end{eqnarray}

\subsubsection{ VPA followed by VPP}

The ratio of decay rates for the
complete VPA-A   followed by VPP decay - exhibited in Fig. \eqref{fig:VPAsequencial} -
can be written as:
\begin{eqnarray} \label{VAP-VPP}
\mbox{Ratio} \;  (VPA-VPP J)  \;  : \;   
\frac{ \Gamma_{\pi^+}^J  }{ 
\Gamma_{\pi^-}^J 
}
&\sim& 
\frac{ 
\left|
\left[ + G_{A_J}^d 
+ G_{A_J}^f D_{A_3} (P^2)  \right] 
D_\rho (P_\rho^2) \left[
+ G_{V^d}     D_{\eta} (P^2_\eta)  - G_{V^f}  
\right]
\right|^2
}{
\left|
\left[ 
+ G_{A_J}^d 
- G_{A_J}^f D_{A_3} (P^2) 
 \right]  D_\rho (P_\rho^2)  \left[
+ G_{V^d}     D_{\eta} (P^2_\eta)  + G_{V^f}  
\right]
\right|^2},
\end{eqnarray}
where $J= 0,8$, $D_\rho (P_\rho^2)$ and $D_\eta (P_\eta^2)$ are respectively 
the rho and eta propagators with corresponding momenta according to 
Figs. \eqref{fig:VPAsequencial} and
 \eqref{fig:VPPdia}.

\section{Numerical estimations }
\label{sec:numerics}
 
An effective transverse gluon propagator, extracted from 
Schwinger Dyson equations at the rainbow ladder approximation,
 was taken from Refs.   \cite{gluonprop-SD} and
 used to compute the meson coupling constants:  
\begin{eqnarray} \label{gluonprop}
R_T (k) &=& 
\frac{8  \pi^2}{\omega^4} De^{-k^2/\omega^2}
+ \frac{8 \pi^2 \gamma_m E(k^2)}{ \ln
 \left[ \tau + ( 1 + k^2/\Lambda^2_{QCD} )^2 
\right]}
,
\end{eqnarray} 
with 
the following parameters 
$\gamma_m=12/(33-2N_f)$, $N_f=4$, $\Lambda_{QCD}=0.234$GeV,
$\tau=e^2-1$, $E(k^2)=[ 1- exp(-k^2/[4m_t^2])/k^2$, $m_t=0.5 GeV$,
$\omega = 0.5$GeV and $D= (0.55)^3/\omega$ (GeV$^2$).
It takes into account a  quark-gluon running coupling constant
and it was extracted from calculations with
Schwinger Dyson equations at the rainbow ladder level.

The parameters used to perform numerical calculations
are those from
  calculations for the 
(pseudoscalar) meson spectrum 
 defined in \cite{PRD-2021}.
 The set of parameters
$X-20$ - $D_{I,2}$ of that work,
calculated with the same gluon propagator of this work,
 provides the following values for the quark effective masses:
\begin{eqnarray} \label{Effmasses}
M_u =0.392 \; GeV,
\;\;\;\;\; M_d =  0.396 \; GeV, 
\;\;\;\;\;     M_s  =  0.600 \; GeV ,
\end{eqnarray}
for $\Lambda = 0.675$ Gev.
These parameters are different from the parameters used in 
\cite{FLB-2024b}, and consequently resulting three-meson
 coupling constants are different (stronger).
Concerning the mixing parameters $G_{i \neq j}$ (for $i,j=0,3,8$),
whose overall
normalizations
are not unambiguous \cite{PRD-2021,JPG-2022}, 
 it is interesting 
to note they are directly proportional to (effective) quark mass  differences.
As a consequence, the hierarchies of mixings in all the   channels
pseudoscalar, scalar, vector and axial mesons,
are  usually  the same,  $G_{08} > G_{38} > G_{03}$,
and this goes  along  with well known results 
\cite{omega-phi}.
To keep the normalization procedure consistent with the effective masses shown above,
the  following meson mixing parameters   -
for the same  set of parameters $X-20$ - $D_{I,2}$  of the quoted work -
were adopted:
\begin{eqnarray}
&& G_{33}^R = 10.00 \; GeV^{-2},
\;\;\;\;\;
G_{88}^R = 7.61  \; GeV^{-2},
\;\;\;\;\;
G_{00}^R =  8.60  \; GeV^{-2},
\nonumber
\\
&& G_{03}^R 
= 0.04 \; GeV^{-2},
\;\;\;\;\;\;
G_{08}^R 
= 1.80 \; GeV^{-2},
\;\;\;\;\;\;
G_{38}^R 
 = 0.05 \; GeV^{-2}.
\end{eqnarray}
It follows  from Eq. \eqref{Gmixbar} that:
\begin{eqnarray} \label{Gmixing}
\bar{G}_{03} =    0.0011
,
\;\;\;\;\;\;
\bar{G}_{08} =   0.0555
,
\;\;\;\;\;\;
\bar{G}_{38} =   0.0014
.
\end{eqnarray}
There are corrections to these values from the sixth order quark interactions
\cite{FLB-2024b}
which are usually  smaller and will  not be taken into account for the present estimations. 

The resulting values of the three-meson coupling constants used in this work 
are the following:
\begin{eqnarray} \label{3mesonscc}
G_{sb,1}^R = -0.216 \; GeV,
\;\;\;\;\;\;
G_{d1}^R = -0.551, 
\;\;\;\;\;
G_{d2}^R = -0.150,
\;\;\;\;\;
G_{sva,sb}^R = 0.018 \; GeV^{-1}.
\end{eqnarray}
The corresponding  couplings, with sometimes different definitions of their coupling constants,
have already been  considered in the literature.
In  \cite{du-zhao} 
the VPA coupling constant, for $\pi-\rho-a_1$,
was found to be:
$G_{v-a-\pi}   \sim 2 M \sim  0.7  \; \mbox{GeV}$ or $G_{VPA} = 2.04 $GeV,
 that are larger  than $G_{sb1}^R$.
A similar estimate   to the present work  was done in
\cite{PRD-2022b},  for a different (re)normalization, that is
 $G_{A\pi V} \sim 0.7- 1.2$.
In \cite{du-zhao} different values were considered for the different channels ASP
with modulus of the order of $1-3$ GeV 
that are  also larger than the values for $G_{d1,2}$.
 always lead to decreasing behavior 
of  each of the couplings 
at higher energies.
For the vertex $\rho-\pi-\pi$, one can define
$G_{\rho-\pi-\pi} = 3 (G_{d1}^R \pm  G_{d2}^R)$.
that are basically  of the
order of magnitude of   values presented 
in the literature, for example as quoted in \cite{PDG}:
$G_{\rho-\pi\pi} = \sqrt{ 4 \; \pi \; 2.9 } \sim 6$.

\begin{figure}[ht!]
\centering
\includegraphics[width=110mm]{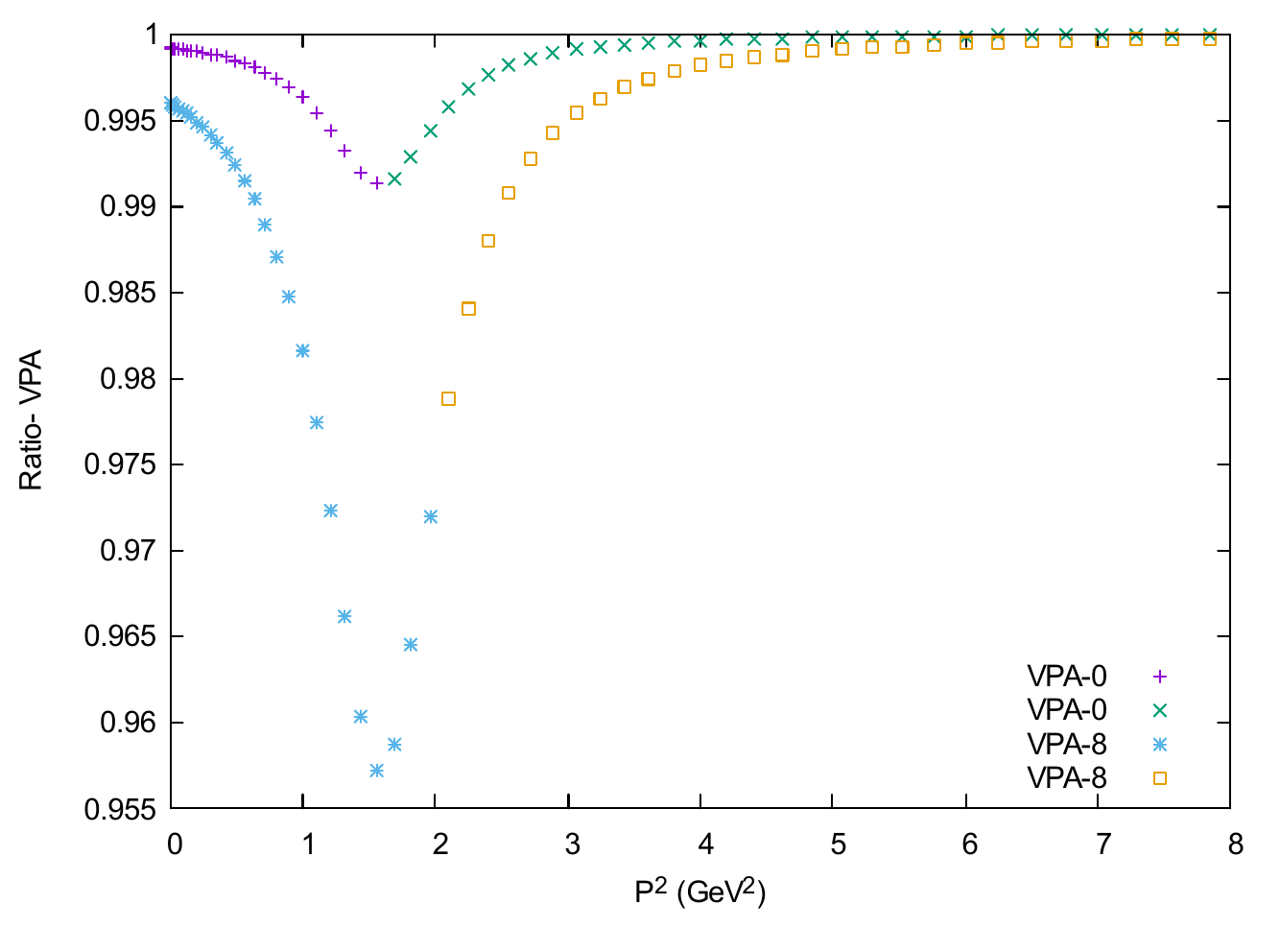}
\caption{ \label{fig:VPAV}
\small
Ratio of decay rates of Eq. \eqref{VAPV-pi+} (VPA)
for $f_1(1285)$ and $f_{1S}(1420)$ (J=0, 8).
 }
\end{figure}

In Fig,  \eqref{fig:VPAV}
the Ratios  VPA-A, Eq. \eqref{VAPV-pi+}, for $J=0$  and $J=8$
are shown as function of the energy $P^2$.
There are  interference effects
that drive the ratio of decay rates $\Gamma_{\pi^+}/\Gamma_{\pi^-}$
 to values  smaller than one,
i.e. destructive interference for positive pion production and 
constructive interference for negative pion production.
This is rather noted in  the  range of energies around the resonance
mass 
of the intermediary states $a_1 (1260)$ induced by meson mixing,
being
$M_{A_1}^2 \simeq 1.6$ GeV$^2$.
The symbols for the ratio are different below and above the 
physical threshold $P^2 = M_{A_0}^2$ (where $A=f_1(1285)$ and $f_{1S}(1420)$).
Maybe there is a way to verify experimentally
 if there is a threshold (and which would be)
for the corresponding different production 
rates of positive and negative pions.
The non-zero 
 width of the intermediary  state
also plays  important roles in the resulting behavior.
The ratio tends to one for energies 
larger than $\sim 4 - 5$ GeV$^2$.
The production of positive and negative rhos, however, has the opposite asymmetry
and it contains an excess of positive rhos in these VPA channel.
Since the
 preferential decay of such positive and negative 
rhos is $\rho^\pm \to \pi^0 \pi^\pm$,  
one may expect that these decays from $\rho^\pm$ tend to restore the ratio 
$\Gamma_{\pi^+}/\Gamma_{\pi^-} = 1$ in this mechanism.
However, 
it may happen that other processes,  such as inelastic scattering or absorption of charged 
rhos,
reduce the 
number of charged rhos available in the (strongly interacting) final state.
Consequently, it is possible to expect that  the ratio  of 
charged pions from the VPA
channel 
of  axial meson decays is different from  the one shown the figure \eqref{fig:VPAV}.
If positive rhos are suppressed in this final state,   more than
negative rhos, this ratio would reach still smaller values.
However,
there is a scenario in which   negative rhos are absorbed more often than positive rhos,
in which  case the ratio of  positive to negative produced pions   
would be larger than one, in this channel.
Another possibility, envisaged below, is that the 
charged rho decay also presents interference effects.

\FloatBarrier

\begin{figure}[ht!]
\centering
\includegraphics[width=110mm]{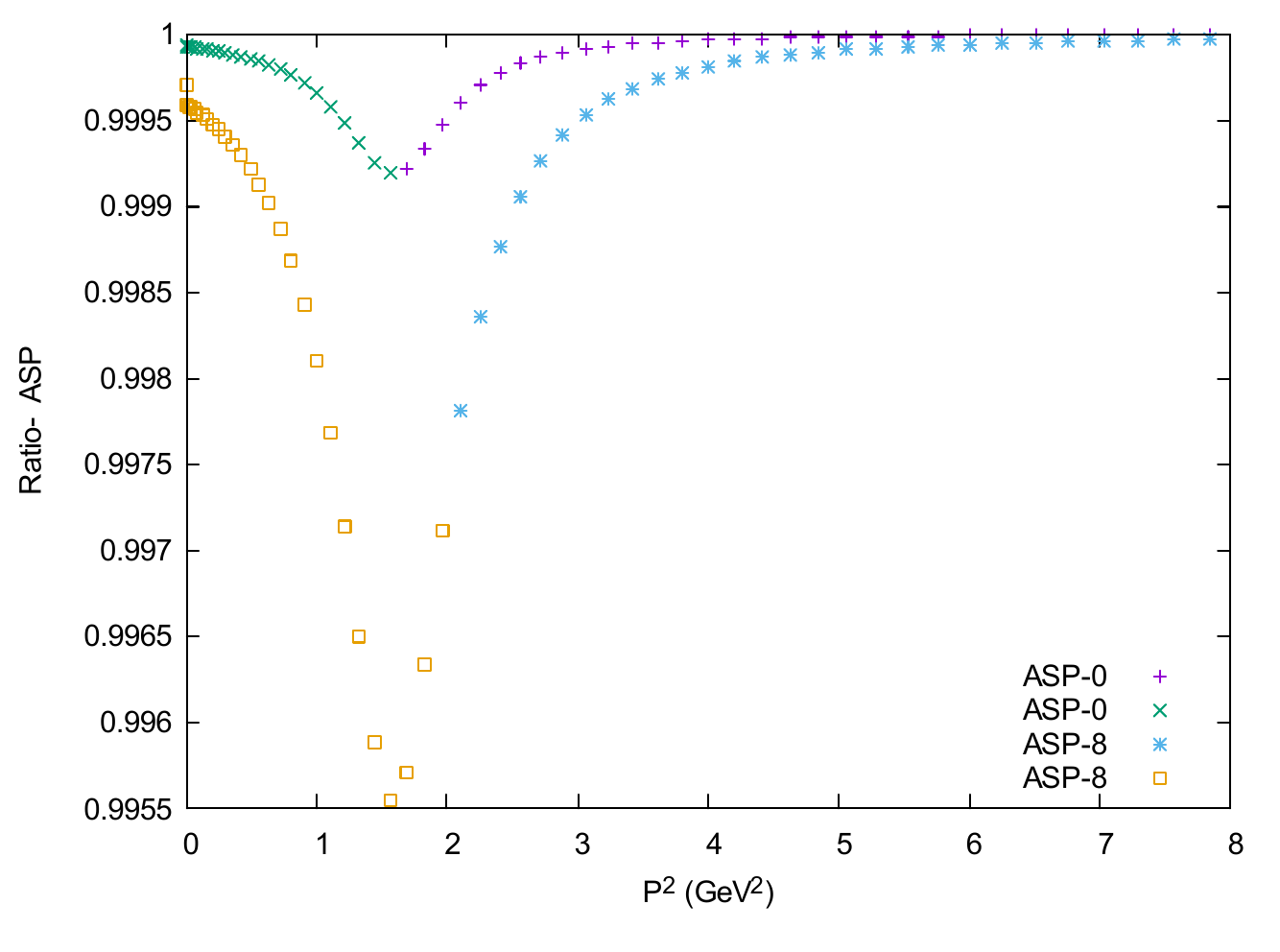}
\caption{ \label{fig:ASPA}
\small
Ratio of decay ratios of Eq. \eqref{ASP-A-pi+}  (ASP-A) processes.
 }
\end{figure}

The ratio \eqref{ASP-A-pi+}  for the  decay with interaction ASP
is exhibited for $J=0, 8$ 
in Fig. \eqref{fig:ASPA}.
The ratio has the same  behavior as  the VPA channel shown above
with the same
 range of energies,  exhibiting destructive (constructive) interference for production of 
positive  (negative) pions.
The interference strength however is smaller due to the 
relative values of the involved coupling constants.
The behavior with energy is the same and the ratio go to one around 
$P^2=3.5$ GeV$^2$ or $P^2=5.0$ GeV$^2$ depending on the channel.
A similar effect to one discussed for  the channel VPA may arise due to the fact that
$a_0^\pm$ may decay into $\pi^\pm \eta$ \cite{PDG}
  and this could restore the same 
number of positive and negative pions in the final state.
However, these charged scalar mesons have a width smaller than the 
width of the rho meson  and 
they may be expected to undergo different effects or scattering
in the  final state   that would prevent them to decay into charged pions.
The final difference between the number of positive and negative pions,
from this mechanism, may
 be therefore smaller than the one indicated in Fig.  \eqref{fig:ASPA}.
Furthermore, the energy dependencies of the couplings - Eqs. \eqref{dPSVAa} and
\eqref{6d-SSV1-mes-2}  -
lead to decreasing functions of the energy
and these effects may lead to smaller asymmetry in the ratio.
Similarly to the VPA channel discussed above, several scenarios are possible.

\FloatBarrier

\begin{figure}[ht!]
\centering
\includegraphics[width=110mm]{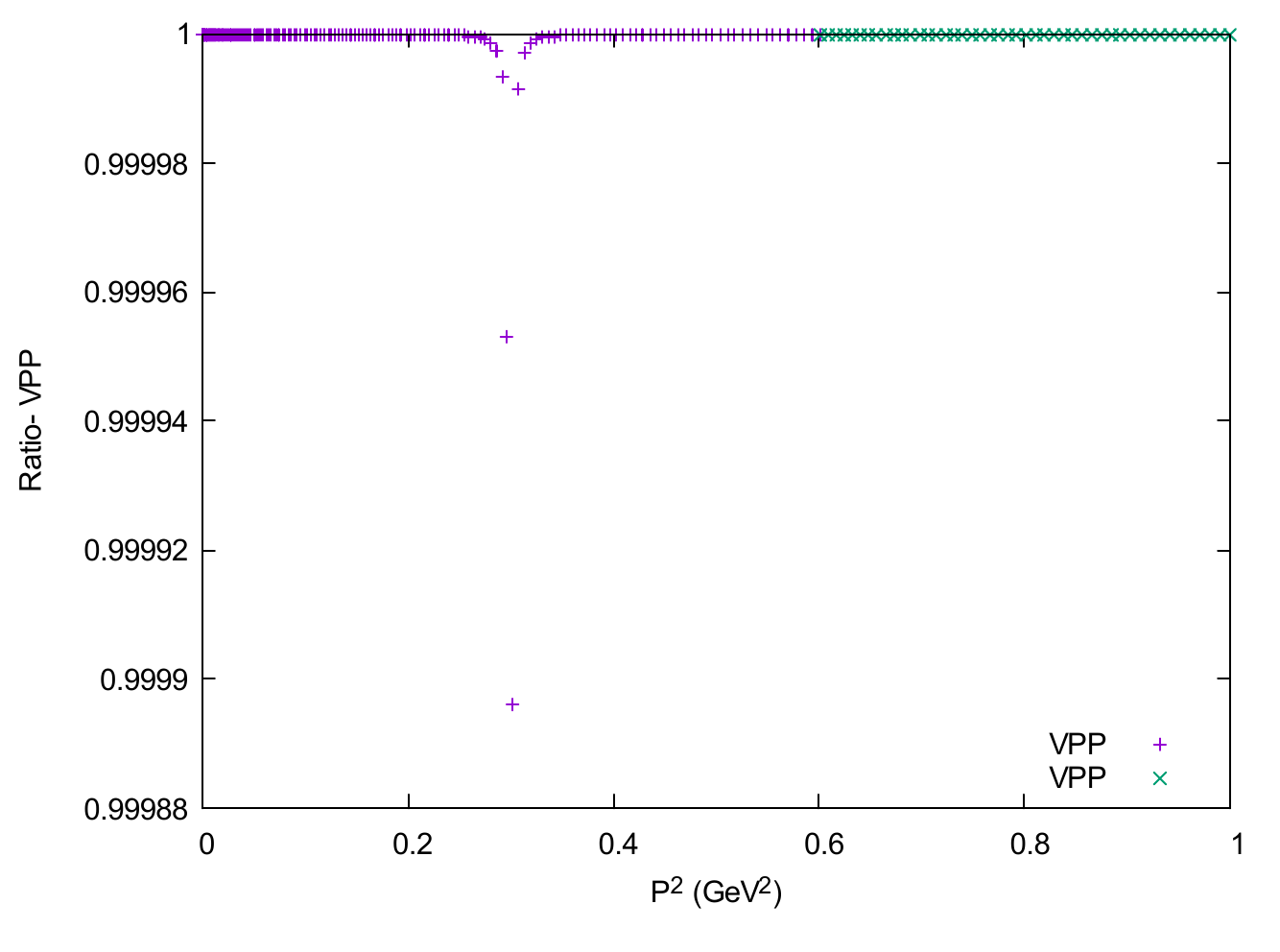}
\caption{ \label{fig:VPP}
\small
Ratio of decay ratios of Eq. \eqref{VPP-ratio}  (VPP)
for $P^2 \equiv P_\eta^2$, different symbols 
above and below the threshold of the rho rest energy $P^2 = M_\rho^2$.
 }
\end{figure}

In Fig. \eqref{fig:VPP}
the ratio \eqref{VPP-ratio}
is presented as a function of the energy  $P^2 \equiv P^2_\eta$,
by using different symbols above and below the 
rho rest energy.
The maximum interference takes place around $P^2 \sim M_\eta^2$
being the width of the $\eta$ very small.
The small deviation from 1 is due to 
the relative values of the coupling constants.
The complete energy dependence of the interactions
should make these values still smaller.
By considering the initial state  $\rho^\pm$  in its rest frame,
the following  CP breaking parameter  have been  calculated 
\begin{eqnarray} \label{CPbreak}
A^\rho &=& 
\frac{ \Gamma_{\pi^+}^{V} - \Gamma_{\pi^-}^{V} }{ 
\Gamma_{\pi^+}^{V} + \Gamma_{\pi^-}^{V} },
\end{eqnarray}
where 
\begin{eqnarray}
\Gamma_{\rho^\pm}^V &=&
\frac{ |M_{VPP,\pi^\pm}|^2  |\vec{K}_\pi | }{ 8 \pi M_\rho^2},
\end{eqnarray}
where 
$|\vec{K}_\pi|$
s the 
 pion momentum and
$M_{VPP}$  is the amplitude for the corresponding channel written in 
Eqs. \eqref{GVd}, \eqref{GVf} and \eqref{VPP-ratio}.  
The charged pion momentum is not the same in the d-channel (where
it shares energy with the $\eta$ from the VPP decay)
 and in the f-channel.
The corresponding pion momentum was considered for each of the 
amplitudes of the d-channel and f-channel, being that,
in the interference term of 
$|M_{VPP,\pi^\pm}|^2$, an average value was taken.
For the charged rho decays
the resulting value is:
\begin{eqnarray}
A^\rho \simeq 2.5 \times 10^{-4},
\end{eqnarray}
This parameter is quite small and it might be reduced 
by the momentum and energy dependence of the interactions or
if the $\eta-\pi^0$ mixing is suppressed for some reason.
It is interesting to emphasize however, that the interference takes place
below the threshold of $M_\rho$ as seen in Fig. \eqref{fig:VPP}.

\FloatBarrier

\begin{figure}[ht!]
\centering
\includegraphics[width=110mm]{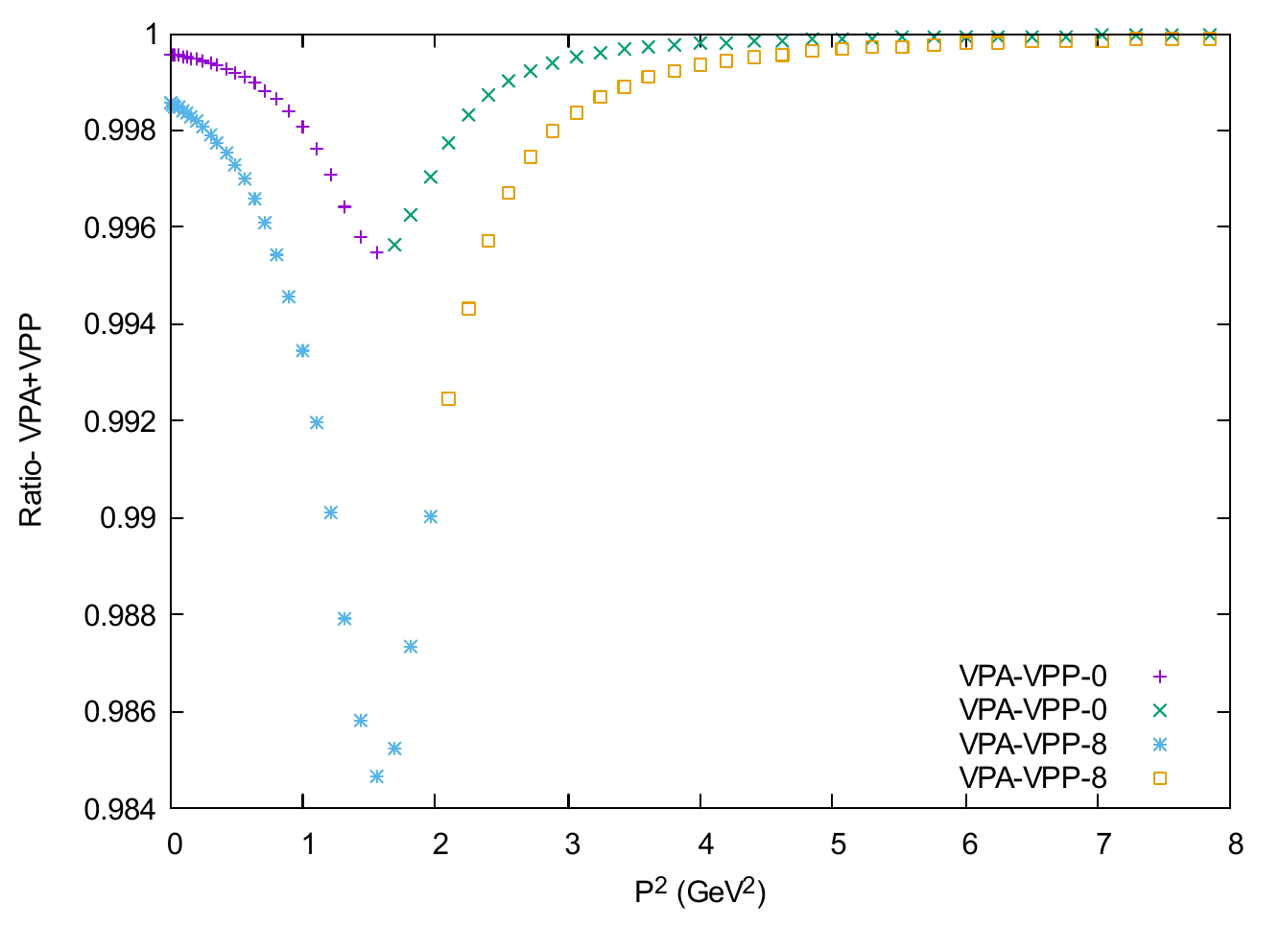}
\caption{ \label{fig:VPA-VPP}
\small
Ratio of decay ratios of Eq. \eqref{VAP-VPP}  for VPA-A process with 
the second decay VPP for both cases, decays of $f_1(1285)$ (J=0) and 
of $f_{1S}(1420)$ (J=8).
 }
\end{figure}

The ratio of the decay  rate \eqref{VAP-VPP} for the VPA interaction with subsequent
VPP decay of the rho meson, by including the possible $\eta-\pi^0$
that leads to interference, as shown in Fig. \eqref{fig:VPP},
is exhibited in Fig. \eqref{fig:VPA-VPP}.
In this case, the d-channel and the f-channel of the VPP decay
are possible in both d- and f-channels of the VPA decay.
Different symbols are used for energies below and above the threshold of each case,
 $f_1(1285)$ and 
  $f_{1S}(1420)$.
The resulting asymmetry is smaller than in the VPA case without VPP decay presented 
in Fig. \eqref{fig:VPAV}.

\FloatBarrier

\begin{figure}[ht!]
\centering
\includegraphics[width=110mm]{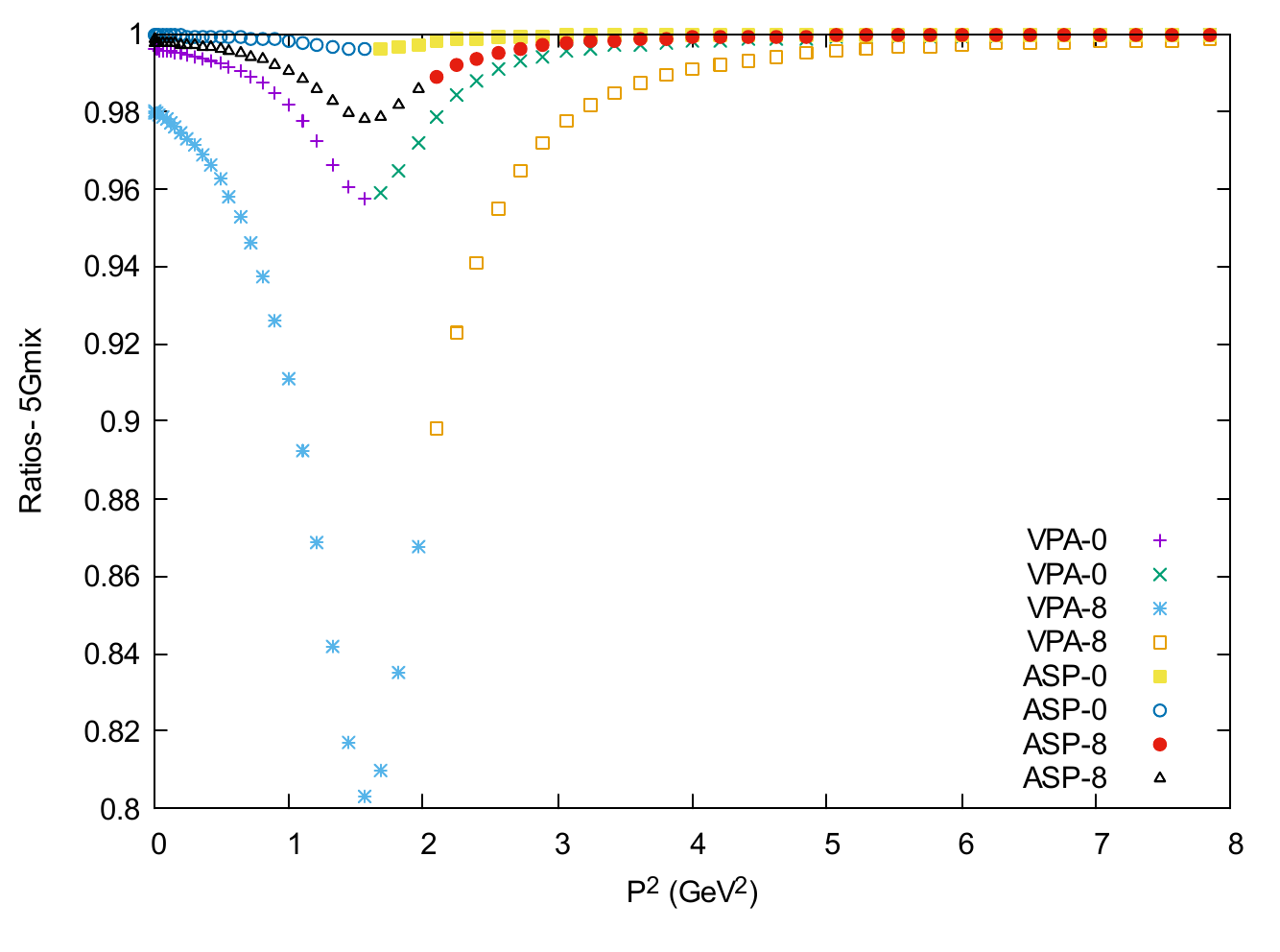}
\caption{ \label{fig:5mix}
\small
Ratios of decay ratios of decay to positive to negative pions for
VPA  and ASP  - 0 and 8 -
 processes,
by considering  mixing interactions  five times larger than the ones  used in the previous figures.
 }
\end{figure}

As discussed above, and in \cite{PRD-2021},
the normalization of the meson mixing parameters  are not unambiguous.
In Fig. \eqref{fig:5mix}
the  ratios  VPA and ASP,  Eqs. \eqref{VAPV-pi+} and \eqref{ASP-A-pi+},
are presented by considering,  arbitrarily,  five times   larger  mixing interactions
than those used above:
$5 \times \bar{G}_{i \neq j}$  for $i,j = 0,3, 8$.
Whereas the behavior of the ratio remains the same,
by comparing with results of the figures above,
 it is seen that 
the interference effect 
 is amplified proportionally to the 
strength of the mixing interactions.

\FloatBarrier

\begin{figure}[ht!]
\centering
\includegraphics[width=110mm]{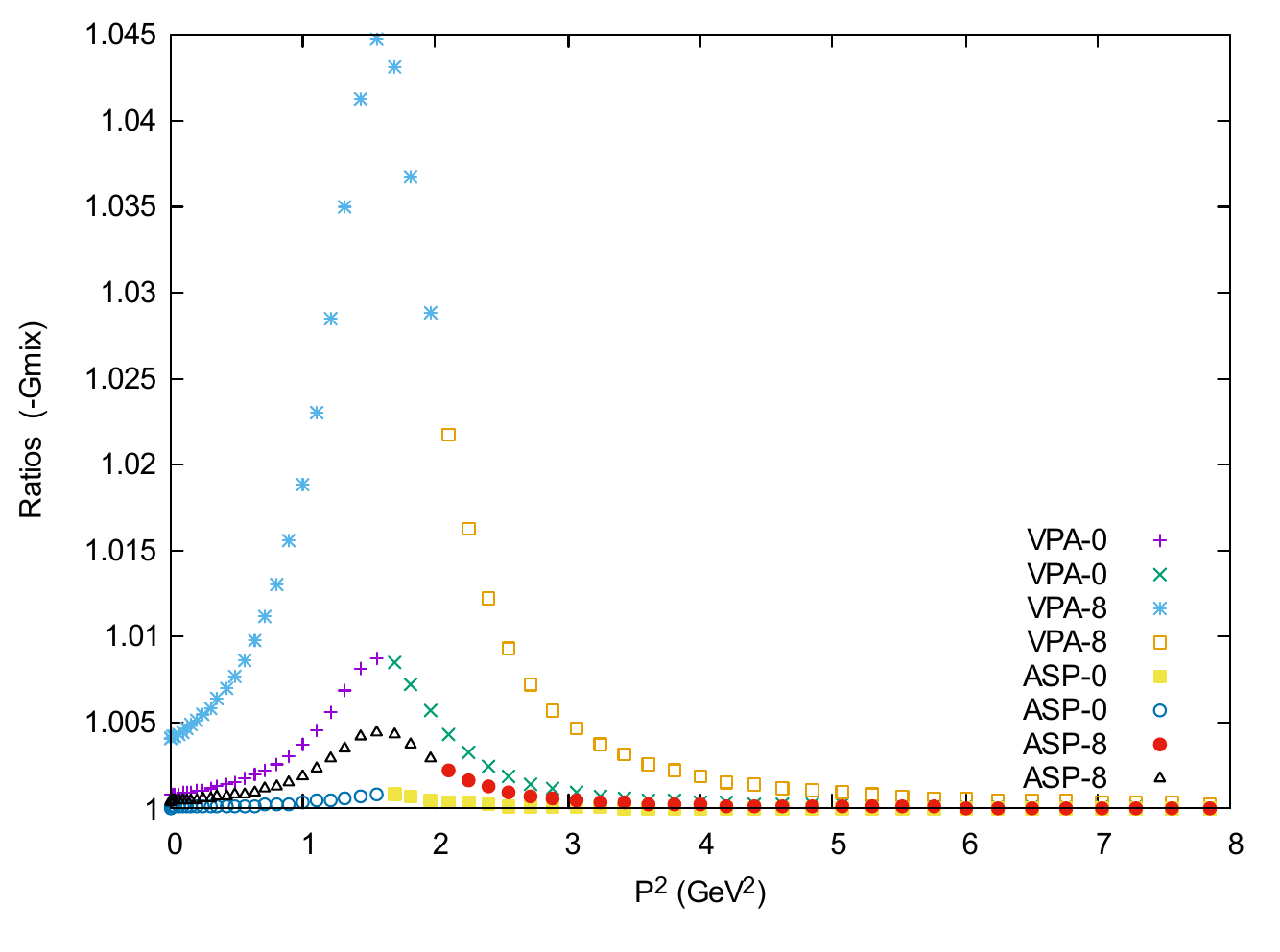}
\caption{ \label{fig:minusmix}
\small
Ratios of decay ratios of decay to positive and negative pions, for
VAP and ASP  with J= 0 and 8
 processes,   Eqs. \eqref{VAPV-pi+} and \eqref{ASP-A-pi+},
by considering  mixing interactions  with signal minus, $- \bar{G}_{i \neq j}$ for i,j=0.3,8.
 }
\end{figure}

In Fig. \eqref{fig:minusmix}
the ratios of  Eqs. \eqref{VAPV-pi+} and \eqref{ASP-A-pi+},
 for VAP and ASP channels (with $J=0$ and $8$), are exhibited for 
the mixing interactions with opposite sign.
The constructive and destructive  interference in the production of 
negative and positive pions are exchanged.

\FloatBarrier

\section{ Summary }

Decays of two  light neutral  axial mesons, $f_1(1285)$ and $f_{1S}(1420)$,
into final states with charged pions   were investigated in this work:
$\; A \to \rho^\mp  + \pi^\pm$ (VPA), 
and  $\; A \to a_0^\mp + \pi^\pm$  (ASP).
These 
processes are energetically favorable contrarily to other possible channels
of the same three-meson interactions.
The  $f_1(1285)$ decays  considered in this work have already been    observed
experimentally
\cite{PDG} and the decays of $f_{1S}(1420)$  have not.
The reasons for this lack of  experimental observation are not clear,
being possibly related to
 very small branching ratio, difficulties with final state interactions and decays 
or even that its structure does not correspond exactly to the state $A_8$ of the
quark-antiquark nonet.
However,  the state  $A_8$ of the possible axial meson nonet can undergo strong mixings
with the states $A_0$ and $A_3$, as discussed in Section  \eqref{sec:mixing},
and these mixings can prevent a clear identification of its structure at the 
example of the $\eta-\eta'$ mixing of the pseudoscalar meson multiplet.
Therefore, 
this investigated was done as a possible scenario.
 The resulting asymmetry in  positive and negative produced
 pion, due to interference, 
is  small, and it is 
directly proportional to the strength of the neutral 
meson mixing parameters for $f_1(1285) - a_1(1260)$
and $f_{1S}(1420)- a_1(1260)$.
The maximum of the interference occurs  
below the threshold energy for the on shell axial meson, either $f_1(1285)$ or 
$f_{1S}(1420)$, 
around the rest energy of the intermediary state $a_1(1260)$, $M_{a_1}$.
 Experimental results  concerning axial  meson mixings 
may be very difficult to be 
obtained  and they  are missing.
This mechanism of asymmetry 
 in the production of  
charged  pions
 due to interference is   suppressed at high energies.
There are at least two effects
that might reduce   further  these values,  or could even change their signs in some scenarios.
The complete  resulting momentum dependencies of the 
used coupling constants are decreasing functions for higher energies,
and this decrease is faster than the possible energy dependencies of the 
interactions in   \eqref{6d-ASP-mes} and \eqref{ddVPA}.
Secondly, 
 the subsequent decay of   charged rhos or scalars $a_0^0$ 
into charged pions may reduce these values 
if these charged particles ($\rho^\pm$ and $a_0^\pm$) do not undergo other
inelastic process that prevents than to decay into a charged pion.
 Subsequent final state interaction with absorption or inelastic scattering 
of $\rho^\pm$ and/or $a_0^\pm(980)$ would play an important role in the resulting 
ratios.
Therefore, further investigation is needed.

The decay $\rho^\pm \to \pi^\pm \pi^0$ was also investigated with 
corresponding three-meson vertices.
The interference is relatively smaller  than in the decay of the axial mesons
and, for the interactions used,
it is  only possible if the mixing $\eta-\pi^0$ is considered.
This mixing may be tested in 
decays of $\rho^\pm$.
However, for the present mechanism,
 the mixing happens below the threshold of the rho rest energy.
A CP-breaking parameter for this decay was estimated to be
 $A^\rho \simeq 2.5 \times 10^{ -4}$.  
Information from experimental observation seem missing.
This decay was also considered for the amplitude of the VPA process, in which 
the decay of the resulting rho also presents interference.
There is a slight decrease in the overall interference,
and therefore in  the ratio $\Gamma_{\pi^+}/\Gamma_{\pi^-}$.

Therefore, the  interference mechanism of asymmetry
 in charged pion production
can be related  to  controversial problems:
neutral meson mixings;  scalar and axial meson structures; and the behavior 
of the final states, i.e, the possibility of different absorption rates or mechanisms for charged
  $\rho^\pm$  and $a_0^\pm (980)$.
It was assumed that the 
 scalar and axial mesons fit, at least in part,  into a  flavor U(3) nonet
although this is not an uncontroversial subject \cite{pelaez,PDG,lightAxial,du-zhao}.
Several other types of interactions
among pseudoscalar, scalar, axial and vector mesons/states,
involving different meson states with strangeness, within this
kind of processes, were left outside the scope of the work.
In addition to that,
a more complete account of  energy dependent effects 
with more complete  
energy and momentum dependencies are currently investigated
and will be  presented in another work.

 \vspace{0.5cm}

\centerline{\bf Acknowledgements}

F.L.B. is a member of
INCT-FNA,  Proc. 464898/2014-5.
F.L.B. acknowledges partial support from 
 CNPq-312750/2021 and CNPq-407162/2023-2.

\end{document}